\documentclass[twocolumn,floatfix, nofootinbib,prd,aps,preprintnumbers,showpacs,showkeys,superscriptaddress,longbibliography]{revtex4-1}

\usepackage[mathscr]{euscript}
\usepackage{amsmath}
\usepackage{graphicx}
\usepackage{dcolumn}
\usepackage{bm}
\usepackage{epsfig}
\usepackage{amssymb,latexsym,mathrsfs}
\usepackage{graphicx}
\usepackage{color}
\usepackage{hyperref}
\usepackage{float}
\usepackage{diagbox}

\usepackage{tikz}

\usepackage{amssymb}
\usepackage{pifont}%

\usepackage{amsmath}
\usepackage{physics}
\usepackage[style=english]{csquotes} 

\usepackage{lipsum, babel}
\usepackage{amssymb}
\usepackage{color}
\usepackage{hyperref}
\usepackage{bm}
\usepackage{amsfonts}
\usepackage{mathrsfs}
\usepackage{graphicx}
\usepackage{amsmath}
\usepackage{float}
\usepackage{braket}
\usepackage{natbib}

     \hypersetup{
    colorlinks=true,
    linkcolor=red,
    citecolor=blue,
} 

\usepackage[caption=false]{subfig}

\usepackage{amssymb}
\usepackage[mathscr]{euscript}
\usepackage{amsmath}
\usepackage{graphicx}
\usepackage{dcolumn}
\usepackage{bm}
\usepackage{epsfig}
\usepackage{amssymb,latexsym,mathrsfs}
\usepackage{graphicx}
\usepackage{color}
\usepackage{hyperref}
\usepackage{float}
\usepackage[thinlines]{easytable}
\usepackage{multirow}
\usepackage{diagbox}
\usepackage{varwidth}
\usepackage{array}
\newcolumntype{P}[1]{>{\centering\arraybackslash}p{#1}}

\newcolumntype{M}[1]{>{\centering\arraybackslash}m{#1}}
\newcolumntype{N}{@{}m{0pt}@{}}

\usepackage{amsmath}

\usepackage{amssymb}
\usepackage{pifont}
\hypersetup{
    colorlinks=true,
    linkcolor=red,
    citecolor=blue,
}

\usepackage{amsmath}
\usepackage{amssymb}
\usepackage[caption=false]{subfig}
\usepackage{hyperref}
\usepackage{url}
\usepackage{xcolor}
\usepackage{color}
\definecolor{amaranth}{rgb}{0.9, 0.17, 0.31}
\definecolor{purple(munsell)}{rgb}{0.62, 0.0, 0.77}
\definecolor{americanrose}{rgb}{1.0, 0.01, 0.24}
\definecolor{palatinateblue}{rgb}{0.15, 0.23, 0.89}
\definecolor{royalblue(web)}{rgb}{0.25, 0.41, 0.88}
\definecolor{hanpurple}{rgb}{0.32, 0.09, 0.98}
\definecolor{beaublue}{rgb}{0.74, 0.83, 0.9}
\definecolor{carminered}{rgb}{1.0, 0.0, 0.22}
\definecolor{brightpink}{rgb}{1.0, 0.0, 0.5}
\definecolor{vividviolet}{rgb}{0.62, 0.0, 1.0}
\definecolor{Uniblue}{RGB}{0,4,100}
\definecolor{crimson}{RGB}{180,0,20}

\definecolor{electron}{rgb}{1.0, 0.67, 0.22}
\hypersetup{ linktoc=all,
    colorlinks, linkcolor={Uniblue},
    citecolor={brightpink}, urlcolor={brightpink}}

\usepackage{physics}
\usepackage{csquotes}
\usepackage{mathtools}
\usepackage{soul}
\makeatletter
\def\l@subsubsection#1#2{}
\makeatother

\begin{document}

\newcommand{\sgn}{\operatorname{sgn}}
\newcommand{\hhat}[1]{\hat {\hat{#1}}}
\newcommand{\pslash}[1]{#1\llap{\sl/}}
\newcommand{\kslash}[1]{\rlap{\sl/}#1}
\newcommand{\lab}[1]{}
\newcommand{\sto}[1]{\begin{center} \textit{#1} \end{center}}
\newcommand{\rf}[1]{{\color{blue}[\textit{#1}]}}
\newcommand{\eml}[1]{#1}
\newcommand{\el}[1]{\label{#1}}
\newcommand{\er}[1]{Eq.\eqref{#1}}
\newcommand{\df}[1]{\textbf{#1}}
\newcommand{\mdf}[1]{\pmb{#1}}
\newcommand{\ft}[1]{\footnote{#1}}
\newcommand{\n}[1]{$#1$}
\newcommand{\fals}[1]{$^\times$ #1}
\newcommand{\new}{{\color{red}$^{NEW}$ }}
\newcommand{\ci}[1]{}
\newcommand{\de}[1]{{\color{green}\underline{#1}}}
\newcommand{\ke}{\rangle}
\newcommand{\br}{\langle}
\newcommand{\lb}{\left(}
\newcommand{\rb}{\right)}
\newcommand{\lbk}{\left[}
\newcommand{\rbk}{\right]}
\newcommand{\blb}{\Big(}
\newcommand{\brb}{\Big)}
\newcommand{\nn}{\nonumber \\}
\newcommand{\p}{\partial}
\newcommand{\pd}[1]{\frac {\partial} {\partial #1}}
\newcommand{\cd}{\nabla}
\newcommand{\cc}{$>$}
\newcommand{\bqa}{\begin{eqnarray}}
\newcommand{\eqa}{\end{eqnarray}}
\newcommand{\bqe}{\begin{equation}}
\newcommand{\eqe}{\end{equation}}
\newcommand{\bay}[1]{\left(\begin{array}{#1}}
\newcommand{\eay}{\end{array}\right)}
\newcommand{\eg}{\textit{e.g.} }
\newcommand{\ie}{\textit{i.e.}, }
\newcommand{\iv}[1]{{#1}^{-1}}

\newcommand{\at}[1]{{\Big|}_{#1}}
\newcommand{\zt}[1]{\texttt{#1}}
\newcommand{\non}{\nonumber}
\newcommand{\m}{\mu}
\def\xa{{m}}
\def\xA{{m}}
\def\xb{{\beta}}
\def\xB{{\Beta}}
\def\xd{{\delta}}
\def\xD{{\Delta}}
\def\xe{{\epsilon}}
\def\xE{{\Epsilon}}
\def\xve{{\varepsilon}}
\def\xg{{\gamma}}
\def\xG{{\Gamma}}
\def\xk{{\kappa}}
\def\xK{{\Kappa}}
\def\xl{{\lambda}}
\def\xL{{\Lambda}}
\def\xo{{\omega}}
\def\xO{{\Omega}}
\def\xvp{{\varphi}}
\def\xs{{\sigma}}
\def\xS{{\Sigma}}
\def\xt{{\theta}}
\def\xvt{{\vartheta}}
\def\xT{{\Theta}}
\def \Tr {{\rm Tr}}
\def\CA{{\cal A}}
\def\CC{{\cal C}}
\def\CD{{\cal D}}
\def\CE{{\cal E}}
\def\CF{{\cal F}}
\def\CH{{\cal H}}
\def\CJ{{\cal J}}
\def\CK{{\cal K}}
\def\CL{{\cal L}}
\def\CM{{\cal M}}
\def\CN{{\cal N}}
\def\CO{{\cal O}}
\def\CP{{\cal P}}
\def\CQ{{\cal Q}}
\def\CR{{\cal R}}
\def\CS{{\cal S}}
\def\CT{{\cal T}}
\def\CV{{\cal V}}
\def\CW{{\cal W}}
\def\CY{{\cal Y}}
\def\BC{\mathbb{C}}
\def\BR{\mathbb{R}}
\def\BZ{\mathbb{Z}}
\def\sA{\mathscr{A}}
\def\sB{\mathscr{B}}
\def\sF{\mathscr{F}}
\def\sG{\mathscr{G}}
\def\sH{\mathscr{H}}
\def\sJ{\mathscr{J}}
\def\sL{\mathscr{L}}
\def\sM{\mathscr{M}}
\def\sN{\mathscr{N}}
\def\sO{\mathscr{O}}
\def\sP{\mathscr{P}}
\def\sR{\mathscr{R}}
\def\sQ{\mathscr{Q}}
\def\sS{\mathscr{S}}
\def\sX{\mathscr{X}}

\def\slz{SL(2,Z)}
\def\slr{$SL(2,R)\times SL(2,R)$ }
\def\ads{${AdS}_5\times {S}^5$ }
\def\adst{${AdS}_3$ }
\def\sun{SU(N)}
\def\ad#1#2{{\frac \delta {\delta\sigma^{#1}} (#2)}}
\def\bqf{\bar Q_{\bar f}}
\def\nf{N_f}
\def\sunf{SU(N_f)}

\def\dcirc{{^\circ_\circ}}

\author{Morgan H. Lynch}
\email{morganlynch1984@gmail.com}
\affiliation{Max-Planck-Institut f\"{u}r Kernphysik, Saupfercheckweg 1, 69117 Heidelberg, Germany}

\title{Hyperbolic recoil and the Unruh effect at CERN-NA63}
\date{\today}

\begin{abstract}
In this manuscript we examine the high energy channeling radiation data sets from the CERN-NA63 experiment using ultra relativistic synchrotron emission. To incorporate recoil, we examine the standard quasi-classical formalism as well as develop a formalism which includes the Unruh effect by utilizing a hyperbolic recoil acceleration, based on conservation of momentum, in the classical synchrotron trajectory. We also perform an asymptotic radiation time scale analysis which predicts a photon energy threshold, beyond which the Unruh effect dominates. We then compare the classical, quasi-classical, and Unruh synchrotron theories to the data. We find that above threshold, the Unruh effect saturates the spectrum of all data sets. 
\end{abstract}


\maketitle
\section{Introduction}
Experimental investigations into the various tenets of quantum field theory in curved spacetime \cite{1967PhDT........75P}, in particular the thermal particle production mechanisms of Parker \cite{1967PhDT........75P, 1968PhRvL..21..562P, 1969PhRv..183.1057P, 1971PhRvD...3..346P}, Hawking \cite{hawking1974black}, Fulling \cite{Fulling:1972md}, Davies \cite{Davies:1974th}, and Unruh \cite{Unruh:1976db}, have been plagued with immense difficulties since their discovery nearly sixty years ago. Particle production by the expanding universe, known as the Parker effect, does have testable predictions in the form of the cosmic microwave background (CMB) trispectrum \cite{Agullo:2010ws}. However, one added difficulty is the fact that there is only one experiment, the universe itself, which can be used to probe the various predictions of the Parker effect. Despite this difficulty, the present thermality of the universe \cite{1976Natur.261...20P}, matter content \cite{1987PhRvD..35.2955F}, and overall structure of the baryon acoustic oscillations do have an interpretation in terms of the Parker effect \cite{2022NatCo..13.2890S}. Particle production by black holes, known as the Hawking effect, is quite difficult to measure due to the fact that the temperatures of astrophysical black holes are vanishingly small, i.e. well below the temperature of the CMB. Intriguingly enough, despite this setback, there is a growing body of evidence that stimulated Hawking emission of gravitational waves \cite{2023PhRvD.108d4047A} via gravitational wave echos \cite{2017PhRvD..96h2004A, abedi2019echoes, 2020Univ....6...43A} could provide a systematic experimental platform to investigate this effect via LIGO. Particle production via acceleration, known as the Fulling-Davies-Unruh effect, has indeed found experimental purchase in the form of high energy particle physics \cite{nico2006observation, bales2016precision, 2018NatCo...9..795W}. It has been found that the radiative beta decay mode of the free neutron is, in fact, a robust manifestation of the Fulling-Davies effect \cite{2023FoPh...53...53G, 2024PTEP.2024b3D01L}, i.e. the thermal dynamical Casimir effect/moving mirror \cite{Davies:1976hi, Davies:1977yv}. Even more intriguing about this system, is the fact that the spectrum is a 1 dimensional Planck distribution. Moreover, the high energy channeling radiation experiments of the CERN-NA63 collaboration have brought the Unruh effect into the realm of experimental science via their explorations of the immense (de)accelerations of quantum radiation reaction \cite{lynch2021experimental, 2024PhRvD.109j5009L, 2025GReGr..57..116L}. It is the continued examination of the NA63 experiments that shall be the focus of the present manuscript.

In the NA63 experiments \cite{2018NatCo...9..795W, 2019PhRvR...1c3014W}, high energy positrons, $E \sim 100$ GeV, are fired into samples of single crystal silicon. These positrons are channeled along a crystalline axis of symmetry, with the positively charged atomic sites forming a transverse confining harmonic potential. Due to the immense Lorentz factor, $\gamma \sim 10^{5}$, the boosted electric fields of the atomic sites brings field strengths sensed by the positrons well into the strong field regime \cite{2012RvMP...84.1177D}. The resulting radiation emission yields photons with energies, $\omega$, on the order of the positron beam energy, $\omega \sim E$. The recoil proper acceleration produced by the photon emission, $a\sim \frac{\omega^2}{m},$ is then sufficiently strong to bring the Unruh effect into the forefront of the dynamics. This fact implies the emitted radiation will have a thermal character, at the Fulling-Davies-Unruh (FDU) temperature, $T_{FDU} = \frac{a \hbar}{2 \pi c k_{B}}$, imprinted on it via a direct interaction with the thermalized quantum vacuum \cite{2017PhRvL.118p1102C, lynch2021experimental}.

Techniques developed to describe the quantum radiation reaction probed by NA63 involve the strong field QED quasi-classical formalism of synchrotron radiation \cite{1998ephe.book.....B,berestetskii1982quantum} as well as techniques which explicitly incorporate the Unruh effect \cite{lynch2021experimental}. Here, we shall focus on the incorporation of the Unruh effect into the strong field QED formalism via the addition of a hyperbolic recoil trajectory in the standard synchrotron trajectory. We will then employ the conventional ultra-relativistic synchrotron theory, the quasi-classical formalism, and the hyperbolic/Unruh recoil formalism to analyze radiation emission measured in the NA63 experiments. This also involves an asymptotic timescale analysis \cite{1986SvPhU..29..788A} which yields a novel condition to determine the energy scale beyond which the Unruh effect saturates the dynamics. The Unruh prescription has been successfully applied to 1 of the NA63 data sets to date, namely the 3.8 mm 178.2 GeV sample \cite{2018NatCo...9..795W}. Now, we will apply the hybrid strong field QED/Unruh prescription to 7 data sets of varying crystal thickness and beam energy \cite{2018NatCo...9..795W, 2019PhRvR...1c3014W} in order to fully establish the presence of the Unruh effect at NA63.

\section{The CERN-NA63 Experiment}

At CERN, the North Area experimental site is fed by the 400 GeV/c proton beam of the super proton synchrotron. Upon hitting the beryllium plate of the T2 target \cite{2019PhRvS..22f1003B}, this proton beam produces showers of secondary beams comprised of numerous particle species. In particular, the H4 beam is then filtered for positrons \cite{Brianti:604383}, and this comprises the source for the NA63 experiments with the specific mandate of examining radiation processes in strong electromagnetic fields \cite{Uggerhoj:881361}. This is accomplished by firing the positron beam into both amorphous and monocrystals of various materials, e.g. silicon, tungston, and gold. A detailed diagram of the beam-sample experimental apparatus is presented in Fig. \ref{na63} below. 

For monocrystals, i.e. single crystal samples with a regular atomic array, the positron beam is channeled, i.e. confined, by the transverse electric field produced by the atomic site nuclei. This transverse electric field is enhanced by a factor of Lorentz gamma, $\gamma$, determined by the beam energy. Thus, as the positron beam passes through the crystal, it will oscillate transversely and radiate. The key feature of the NA63 experiment is that this enables the exploration of radiation processes at, or near, the critical electric field, $E_{cr} = \frac{m_{e}^2}{q} =  1.32 \times 10^{18}$ V/m. As an example, consider the typical Coulomb field produced by the nucleus of an atom, $E_{Z} = \frac{ k Z q}{r^2}$. Here, $k$ is Coulombs constant, $q$ is the elementary electron charge, and $Z$ is the atomic number. These atoms, for monocrystals, will be spaced by a distance characterized by their lattice constant, $\ell_{Z}$. Note, for silicon we have, $Z=14$ and $\ell_{14} = .54$ nm. Thus, for a channeling experiment with a 178.2 GeV beam energy, and an effective distance ranging from $r = \ell_{14} \leftrightarrow \ell_{14}/10$ from the atomic center, the positrons feel a transverse electric field strength ranging from $\frac{\gamma E_{14}}{ E_{cr}} = .02 \leftrightarrow 1.83$. It is these large electric fields which probe the physics near the critical field strength that enables NA63 to explore novel radiative phenomena such as the production of particles from the vacuum and analogues of Hawking radiation from black holes \cite{CERNCourierVolume51:1734657}. A close variant of the latter, namely the Unruh effect, is the subject of the current manuscript.

\begin{figure}[]
\centering  
\includegraphics[scale=.52]{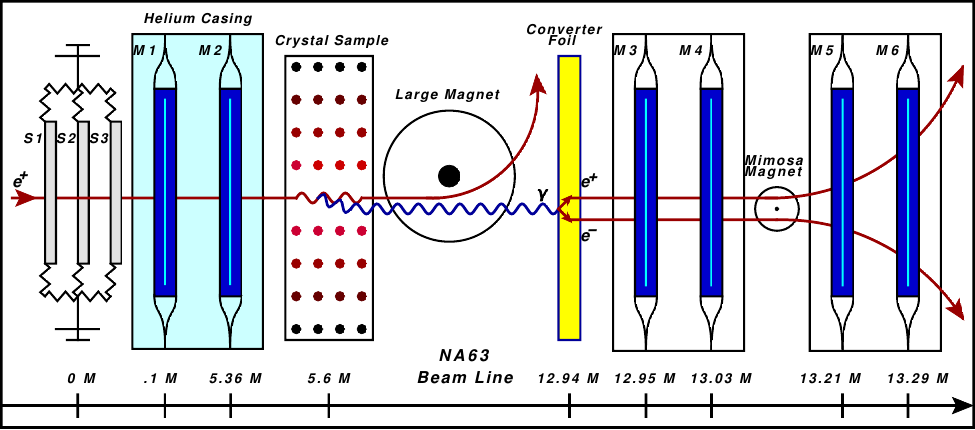}
\caption{A schematic diagram of the NA63 beam line for the 178.2 GeV experimental setup \cite{2018NatCo...9..795W}. Here we have 3 scintillators labeled $Sj$ and 6 position dependent MIMOSA detectors labeled $Mi$. From the left, a positron (red) has its initial position and energy measured and then, upon entering the crystal sample, undergoes a transverse channeling oscillation and emits a photon (blue). A large magnet bends the original positron away and the emitted photon is converted into an electron-positron pair which then passes through a Mimosa magnet. The bend in the electron-positron trajectory is used to reconstruct the photon energy. } 	
\label{na63}
\end{figure}

\section{Incorporating Radiation Reaction}

The electric field strength present at NA63 also simplifies the overall dynamics involved in radiation emission. In channeling radiation, one can typically describe the classical positron trajectory via a longitudinal beam velocity, $v_{b} = v_{0}$, along with a transverse oscillation, $v_{\perp} = -v_{t} \sin{(\Omega t)}$, modulated by the channeling oscillation frequency $\Omega$. This transverse oscillation is produced by the atomic potential of the crystal atomic sites. The frequencies of photons radiated will then be at multiples of the channeling oscillation frequency, $\omega \sim n\Omega$, see \cite{1983ARNPS..33..453A}. In the case of strong field dynamics, the emitted photon will have energies on the same order as the positron beam energy, $\omega \sim E$. Thus we are restricted to single photon emission. This drastically simplifies the problem into that of ultra relativistic synchrotron emission, where the positron trajectory can simply be modeled via, $v_{z} = v_{0} \cos{(\Omega t)}$ and $v_{\perp} = -v_{0} \sin{(\Omega t)}$. In this regime, one then needs only consider dynamics which occur over a timescale such that $\Delta t \Omega \sim 1/\gamma$ \cite{berestetskii1982quantum}.  Moreover, in this strong field regime we have a convergence of high energy channeling \cite{2018NatCo...9..795W}, high energy synchrotron \cite{1968JETP...26..854B}, and high energy electron-laser emission \cite{1979MINFI.111....5R} to similar theoretical descriptions.

The incorporation of recoil, or radiation reaction, in these theories is accomplished in two ways: the quasi-classical method or by the use of exact quantum wavefunctions. In the case of synchrotron/channeling emission, the techniques employed involve the commutation of photon field operators with electron field operators. As a sketch, let us consider an initial electron wave function, $\hat{\psi_{i}} \sim u_{i}(\hat{p})e^{-i H(\hat{p}) t}$, final electron wave function,  $\hat{\psi_{f}} \sim u_{f}(\hat{p})e^{-i H(\hat{p}) t}$, and photon wave function,  $A_{k}\sim e^{-i k x}$. The resultant interaction is given by, $\hat{S} \sim \hat{\psi_{f}} A_{k} \hat{\psi_{i}} $. Then, by commutation of the photon field operator to the left, we can use the wave function itself as an operator for momentum transformations via, $\hat{p}e^{-i k x} = e^{-i k x} (\hat{p} - k) $ and $H(\hat{p}) e^{-i k x} = e^{-i k x} H(\hat{p} - \hat{k}) $. Thus, we are able to incorporate recoil, via the quasi-classical formalism, into the final state electron observables by commutation \cite{1968JETP...26..854B}, e.g. $p_f = p_i -k$. The case with exact quantum wavefunctions is much simpler conceptually, although significantly more difficult mathematically. This involves the use of exact solutions for Dirac particles in a laser field background, i.e. Volkov states. Thus, recoil is simply incorporated by computing the transition amplitudes between initial and final state wavefunctions \cite{1979MINFI.111....5R,1964PhRv..133..705B}.

In this manuscript, we will employ a theory which directly incorporates recoil into the positron trajectory. To this end, we shall consider the synchrotron limit of the channeling trajectory, $v_{z} = v_{0} \cos{(\Omega t)}$ and $v_{\perp} = -v_{0} \sin{(\Omega t)}$ and include a hyperbolic recoil acceleration, $v_{a} =  \frac{at/\gamma}{\sqrt{1 + \lb \frac{at}{\gamma } \rb^{2}}}$, under relativistic velocity addition. Moreover, we will work in the ultra relativistic limit where the photon emission is beamed forward with an emission angle, $\sin{(\theta)} \sim \frac{1}{\gamma}$. In other words, the z-component of the photon momentum is given by $k_{z} = \omega \cos{(\theta)} \approx \omega$ while the transverse component is suppressed via, $k_{\perp} = \omega \sin{(\theta)} \approx \frac{\omega}{\gamma}$. As such, for a lab frame analysis of an ultra relativistic positron, it suffices to incorporate recoil into the z-component of the final velocity via, $v_{f} = \frac{v_z-v_a}{1-v_z v_a}$, i.e. the problem becomes 1 dimensional in the ultra relativistic limit, see section (\ref{1d}). In the hyperbolic trajectory, $v_a$, we shall utilize a recoil acceleration produced by the photon emission itself. For synchrotron trajectories, photon recoil can lead to three types of radiation reaction: inspiral, undulation, and guiding center recoil. A positron propagating along a circular trajectory will radiate its kinetic energy into photons and diminish its radius of curvature leading to an inspiral. These photons radiated will also reside inside a cone of opening angle $\theta \sim \frac{1}{\gamma}$ with emission into directions $\sim \theta /2$ both above and below the direction of propagation, thereby leading to undulation. Finally, when the photon emission is sufficiently energetic, the kick produced by the radiation will cause the entire ``disk" swept out by the circular orbit to shift or translate in the direction opposite the photon emission, i.e. guiding center recoil \cite{1983JPhA...16L.669L, 1984JPhA...17L..91L, 1984JPhA...17L.223L, 1984JPhA...17.2895D, 1985JPhA...18L.111W, 1987JPhA...20.2105P, 1987JPhA...20.2405L, 1987A&A...176L..21L}. As we shall see, it will be this guiding center recoil \cite{1987JPhA...20.2405L, 1987A&A...176L..21L} that is also responsible for the onset of the Unruh effect, see section (\ref{ll}) below for more details. Let us now examine in more detail how the Unruh effect will manifest in the experiments carried out by NA63.

\section{Radiation Time Scales}
One of the key hurdles to overcome in exploring the Unruh effect experimentally is to find a system where the acceleration has a duration long enough for it to saturate the dynamics. Typically this comes in the form of a ``thermalization time" whereby a comoving energy gap, $\Delta E$, has time to reach detailed balance. This implies that given an acceleration, and thus $T_{FDU}$, the excitation rate of the gap, $\Gamma_{\uparrow}$, and the de-excitation rate, $\Gamma_{\downarrow}$, has the standard Boltzmann relation $\Gamma_{\uparrow} = \Gamma_{\downarrow}e^{-\Delta E/T_{FDU}}$. This thermalization time, $\tau_{th}$, is given by the reciprocal of the excitation rate, $\tau_{th} = 1/\Gamma_{\uparrow}$.

It should be noted that the above thermalization time is an absolute time scale for a given energy gap. In a standard experimental setting, there will be multiple processes present which can give rise to radiation emission or transitions in the energy gap. Thus, in order to determine when the Unruh effect will dominate the dynamics, we must compare the relevant time scales in a given system. In this regard, we shall perform an asymptotic radiation analysis which enables us to determine the radiation formation and coherence times of radiative processes, in this case synchrotron/channeling radiation, both with and without a hyperbolic recoil, i.e. the Unruh effect. When the associated time scales of the hyperbolic recoil/Unruh effect becomes smaller than that of synchrotron emission, i.e. it happens faster, we expect the Unruh effect to dominate the dynamics of the system.  

To begin our asymptotic timescale analysis \cite{1986SvPhU..29..788A}, let us consider the action for a general radiative process, e.g. a photon $A^{\m}(x)$ radiated by the charge current $j_{\m}(x)$ using the standard QED interaction,
\bqe
\hat{S}_{I} = q\int d^{4}x j_{\m}(x)\hat{A}^{\m}(x).
\eqe
Here, our charged current is given by $j_{\m}(x) = v_{\m}(t) \delta{ (\mathbf{x} - \mathbf{x}(t))}$, with $v_{\m} = \frac{d x_{\m}}{dt}$ characterizing the charged current trajectory, and we shall use the standard mode decomposition for a massless vector fields,
\bqa
\hat{A}^{\m}(x) &=& \int \frac{d^{3}k}{(2 \pi)^{3/2}} \frac{\sum_{i}\epsilon_{i}^{\ast \m}}{\sqrt{2\omega}} \\ \non
&\times &\lbk \hat{a}_{k}e^{i(\mathbf{k}\cdot \mathbf{x} - \omega t) } + \hat{a}^{\dagger}_{k}e^{-i(\mathbf{k}\cdot \mathbf{x} - \omega t)}  \rbk.
\eqa
We can now formulate the photon emission amplitude, $\mathcal{A} = \bra{\mathbf{k}}\hat{S}_{i} \ket{0}$. For the emission of a photon with a definite momentum, $k$, and polarization, $\epsilon^{\m}$, this photon matrix element yields,
\bqe
\bra{\mathbf{k}} \hat{A}^{\m}(x)\ket{0} = \frac{1}{(2 \pi)^{3/2}} \frac{\epsilon^{ \ast \m}}{\sqrt{2\omega}} e^{-i(\mathbf{k}\cdot \mathbf{x} - \omega t)} .
\eqe
Our total amplitude then yields,
\bqa
\mathcal{A} &=&  \frac{q}{(2 \pi)^{3/2}}\int d^{4}x  \delta{ (\mathbf{x} - \mathbf{x}(t))}
  \frac{v_{\m}(t)  \epsilon^{\ast \m}}{\sqrt{2\omega}} e^{-i(\mathbf{k}\cdot \mathbf{x} - \omega t)} \\ \non 
  &\sim &  \int dt \mathbf{v}(t) e^{-i(\mathbf{k}\cdot \mathbf{x} - \omega t)}. \label{amp}
\eqa
Since both the velocity and oscillatory exponent are smooth functions, we will evaluate the above amplitude via the method of steepest descent \cite{1986SvPhU..29..788A}. Due to the highly oscillatory nature of the integrand, the leading contribution to the integration can be determined via the stationary phase approximation.  Recalling, $\textbf{k}$ and $\omega$ are the photons momentum and frequency respectively, we will consider the relevant photon phase, $\phi =\textbf{k}\cdot\textbf{x} - \omega t$, this means we must seek out solutions for $\frac{d\phi}{dt} = 0$. From this we have, $1-\beta(t) \cos{(\theta)} = 0$, and recalling that the angle of emission $\sin{(\theta)} \sim 1/\gamma$, we will then have $\cos{(\theta)} \sim \beta(t)$. This condition then amounts to
\bqa
1-\beta^2(t_{0})  = 0 \non \\
\Rightarrow t_{0} = t_{1} + i t_{f}.  \label{phase}
\eqa
Here, the velocity, $\beta(t) =v(t)/c$, characterize the semiclassical trajectory of charged current. We should note that Eqn. (\ref{phase}) will, in general, not have any real solution for charged particles moving in free space. We will then consider the complex solutions for our stationary phase. As such, our phase can be written as,
\bqe
\phi(t) =\phi(t_{0})+\frac{d\phi(t_{0})}{dt}(t - t_0)+ \frac{1}{2}\frac{d^2\phi(t_0)}{dt^2}(t - t_0)^2 +\cdots
\eqe
We recall that the second term in the above expansion vanishes by virtue of the stationary phase. Utilizing the above expansion in our amplitude, Eqn. (\ref{amp}), we have
\bqe
\mathcal{A}  \sim  \mathbf{v}(t_0) e^{-i(\mathbf{k}\cdot \mathbf{x}(t_0) - \omega t_0)} \int dt e^{-\frac{i}{2}\mathbf{a}(t_0) \cdot \mathbf{k}(t - t_0)^2 }. \label{amp1}
\eqe
Evaluation of the above standard Gaussian integral will then yield,
\bqe
\mathcal{A}  \sim  \mathbf{v}(t_0)\sqrt{\frac{2 \pi}{ | \mathbf{a}(t_0) \cdot \mathbf{k}|}} e^{-i(\mathbf{k}\cdot \mathbf{x}(t_0) - \omega t_0)} e^{i\phi}. \label{amp2}
\eqe
Here, the angle $\phi$ determines the direction of steepest descent which, upon computation of any relevant observable which can be derived by the absolute value of the amplitude, is only an irrelevant overall phase. As such, it is left arbitrary. Now, the above amplitude does contain an overall smooth prefactor, however the asymptotic form will be controlled by the exponential form in the high frequency limit. In this regard, we note that by using $t_{0} = t_{1} + i t_{f}$, the trajectory in the exponent can be written as, $\mathbf{x}(t_0) = \mathbf{x}_{1}+i\mathbf{x}_f$. As such we have 
\bqe
\mathcal{A} \sim  \mathbf{v}(t_0)\sqrt{\frac{2 \pi}{ | \mathbf{a}(t_0) \cdot \mathbf{k}|}}  e^{i\phi} e^{-i(\mathbf{k}\cdot \mathbf{x}_1 - \omega t_1)} e^{-(\omega t_f -\mathbf{k}\cdot \mathbf{x}_f  )}. \label{amp3}
\eqe 
The above expression then gives us our first insight into the asymptotic form of the radiation emission. Here, we have an overall smooth prefactor, oscillatory phase, and the exponentially decaying term which governs the high frequency tail of the radiation emission. Let us examine the exponential damping term in more detail,
\bqe
\mathcal{A} \sim   e^{-(\omega t_f -\mathbf{k}\cdot \mathbf{x}_f  )}. \label{amp4}
\eqe 
In order to simplify the above expression, let us expand the trajectory about the point $t_f = 0$. Note, this also implies, $dt_0 = dt_1$. Then, our trajectory term can be simplified via the following;
\bqe
\mathbf{x}(t_0) = \mathbf{x}(t_1) + i t_f \mathbf{v}(t_1).
\eqe
Here we have defined, $\mathbf{v}(t_1) = \frac{d \mathbf{x}(t_1)}{dt_1} $. From the above expression, we can determine the imaginary component, $\mathbf{x}_f $, which governs the exponential decay to be $\mathbf{x}_f = t_f \mathbf{v}(t_1)$. Our damping factor then takes the following form,
\bqe
\mathcal{A} \sim   e^{-t_f(\omega  -\mathbf{k}\cdot \mathbf{v}(t_1)  )}. \label{amp5}
\eqe
Note, we can now define the coherence time, $t_{c}$, which determines the overall timescale where the radiation is emitted coherently given by, 
\bqe
t_{c} = \frac{1}{\omega(1-\beta(t_{1})^2)} . \label{tform}
\eqe
The subsequent radiation amplitude will be determined by the ratio of the formation time to the coherence time and will have an asymptotic exponential decay determined by \cite{1986SvPhU..29..788A},
\bqe
\mathcal{A} \sim e^{-t_{f}/t_{c}}.
\eqe
The power spectrum of radiation, $\frac{d\mathcal{P}}{d\omega}$, is then determined by the square of this amplitude. Hence,
\bqe
\frac{d\mathcal{P}}{d\omega} \approx e^{-2t_{f}/t_{c}}. \label{pow}
\eqe
In the above asymptotic analysis there are three time scales that can be extracted from the above condition which determine the asymptotic characteristics of the emitted radiation. From the root of our stationary phase, Eqn. (\ref{phase}), we have $t_1$ and $t_f$. Here, $t_{1}$ determines the time when the radiation is emitted. The ``formation time", $t_{f}$, determines the time interval over which the waves of radiation are built up and emitted with a well defined phase. Finally, the coherence time, $t_c$, determines how long that well defined phase will last. In short, during an experiment we began observing radiation at $t_1$, we measured a radiation pulse comprised of many photons for $t_c$ seconds, and each of the measured photons took $t_f$ seconds to form. We must also note, that for a radiative process to take place, we must have $t_f<t_c$. This way, we ensure that that radiation has time to form within the time scale where the emission is coherent, see Fig. \ref{channel} below. Thus yielding a strong emission prior to the exponential cutoff. In the following sections, we shall examine these relevant time scales and spectra for radiation emitted from a synchrotron-like oscillation, e.g. channeling radiation, and from a synchrotron oscillation which includes a hyperbolic ``recoil" trajectory which encodes the Unruh effect.

\subsection{Radiation formation times}
For a channeling-like oscillation, with frequency $\Omega$ and max velocity $v_{0}$, our velocity profile will be given by,
\bqa
v_{z} &=& v_{0}\cos{(\Omega t)} \non \\
v_{\perp} &=& -v_{0}\sin{(\Omega t)}. \label{chan}
\eqa
If we consider the photon to be emitted forward, along the z-axis, it suffices to only consider the z-component of the channeling trajectory. Now, for proper acceleration, $a$, we will also consider the hyperbolic recoil trajectory in the boosted frame of the particle, $x^2 - t^2/\gamma^2 = 1/a^2$. Here $t$ is the time measured in the lab frame. Determination of the associated recoil velocity yields,  
\bqe
v_{a} =  \frac{at/\gamma}{\sqrt{1 + \lb \frac{at}{\gamma } \rb^{2}}}. \label{accel}
\eqe
We must also keep in mind that the photon emission causes a recoil in the negative z direction. With this recoil in the boosted frame, we must add its velocity to the z-component of the initial channeling velocity of the beam, $v_{z}$ via relativistic velocity addition. We will then have the following lab frame velocity,
\bqe
v(t) =  \frac{v_{z}-v_{a}}{1-v_{z}v_{a}} \label{rv}
\eqe
The phase condition, Eqn.~(\ref{phase}), then yields,
\bqa
1- \lbk \frac{v_{z}-v_{a}}{1-v_{z}v_{a}}  \rbk^2 &=& 0 \non \\
\lbk 1-v^{2}_{a} \rbk \lbk 1-v^{2}_{z} \rbk &=& 0
\eqa
Here we see that the condition which determines our formation times factorizes into the component from each velocity. Let us now examine each case. Beginning with the channeling oscillation, we perform a series expansion in $\Omega t$ until we find our first imaginary solution. Hence,
\bqa
1-\lbk v_{0}\cos{(\Omega t)} \rbk^2 &=& 0\non \\
\frac{1}{\gamma^2} + (\Omega t)^2 &=& 0 \non \\
\Rightarrow t_{0\Omega} = 0 + i\frac{1}{\Omega \gamma}.
\eqa
Here we find, $t_{1\Omega} = 0$ and a formation time of the channeling oscillation of $t_{f\Omega}= \frac{1}{\Omega \gamma}$. The coherence time for the channeling oscillation, Eqn. (\ref{tform}), is then found to be,
\bqe
t_{c\Omega} = \frac{\gamma^2}{\omega}.
\eqe
Next, we shall examine the time scales for the hyperbolic recoil acceleration. Again, performing a series expansion in $\frac{at}{\gamma}$ until we find our first imaginary solution, we find
\bqa
1-\lbk  \frac{-at/\gamma}{\sqrt{1 + \lb \frac{at}{\gamma } \rb^{2}}} \rbk^2 &=& 0\non \\
1- \lb \frac{at}{\gamma } \rb^{2}+\lb \frac{at}{\gamma } \rb^{4}&=& 0 \non \\
\Rightarrow t_{0a} = \frac{\sqrt{3}\gamma}{2a} + i\frac{\gamma}{2a}.
\eqa
Here we have taken the positive root of both components. As such we find, $t_{1a} = \frac{\sqrt{3}\gamma}{2a}$ and a formation time of the hyperbolic recoil of $t_{fa} = \frac{\gamma}{2a}$. The coherence time for the hyperbolic recoil is then found to be,
\bqe
t_{ca} = \frac{7}{4\omega}.
\eqe

\begin{figure}[]
\centering  
\includegraphics[scale=.52]{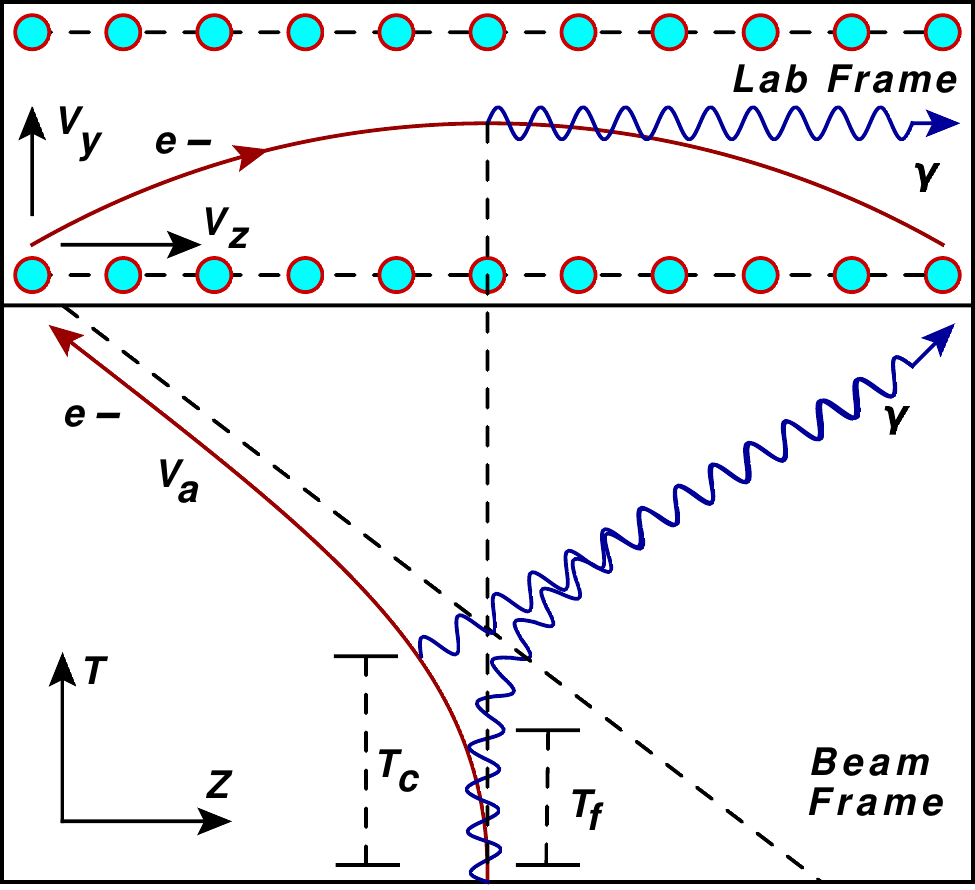}
\caption{A schematic diagram of the positron trajectory and photon emission process for both the lab frame and beam frame. In the lab frame we have a positron undergoing a synchrotron trajectory, produced by the confining electric field of the atomic sites, and radiating. In the beam frame, we have a hyperbolic recoil produced by the photon emission itself. We also see the dynamics for the formation time of each photon, $t_f$, and the overall coherence time for the photon emission, $t_c$.  } 	
\label{channel}
\end{figure}

\subsection{Unruh threshold}

We now have the relevant time scales which comprise our two spectra, one is due to channeling/synchrotron radiation, Eqn.~(\ref{chan}), and the other due to a hyperbolic recoil/Unruh effect, Eqn. (\ref{accel}). To see when the recoil emission dominates over the synchrotron emission, let us now turn to the power spectra analysis. We will look for the scenario where the asymptotic power spectrum due to the recoil acceleration dominates the power spectrum due to the synchrotron radiation, i.e. $\frac{d\mathcal{P}_{a}}{d\omega}  >  \frac{d\mathcal{P}_{\Omega}}{d\omega}$. Thus,
\bqa
e^{-2t_{fa}/t_{ca}} &>& e^{-2t_{f\Omega}/t_{c\Omega}} \non \\
\frac{t_{fa}}{t_{ca}} &<& \frac{t_{f\Omega}}{t_{c\Omega}} \non \\
\frac{2\gamma \omega}{7a} &<& \frac{\omega}{\Omega \gamma^3}.
\eqa 
The above condition has a very simple interpretation; When the radiation time scales produced by the recoil are shorter than that of the synchrotron radiation, the recoil process ``embezzles" the photon from the synchrotron process and therefore dominates the emission. This threshold then sets a condition on the photon energy where the hyperbolic recoil, and thus the Unruh effect, will dominate the spectrum. In this regard, we shall examine a recoil acceleration of the form $a \sim \frac{\omega^2}{m}$ \cite{lynch2021experimental, 2024PhRvD.109j5009L, 2025GReGr..57..116L}. Moreover, the channeling frequency, $\Omega$, can be mapped to the typical emitted photon frequency via, $\omega = 2 \Omega \gamma^3$ \cite{2012RvMP...84.1177D}. Note, here we have included the factor of 2 for later convenience. We then have our critical photon energy, $\omega_c$, given by, 
\bqe
\omega_c > \frac{E}{7}. \label{cond}
\eqe 
For photons frequencies which are above this critical frequency, we expect the Unruh/recoil power spectrum to dominate over the synchrotron spectrum. In other words, for photons which obey Eqn.~(\ref{cond}), the hyperbolic trajectory is the source of the radiation and therefore brings the Unruh effect into the forefront of the particle spectrum. Let us now consider a suite of NA63 channeling radiation experiments where we will have two positron beam energies, 50 and 178.2 GeV \cite{2019PhRvR...1c3014W, 2018NatCo...9..795W}. This means our critical frequencies, above which our hyperbolic recoil dominates, will be given by,
\bqa
\omega_{178.2} &=& 25.5 \; \text{GeV} \non \\
\omega_{50} &=& 7.1 \; \text{GeV}.
\eqa
We must also note the fact that $\omega_{178.2} = 25.5$ GeV is very similar to the thermalization time threshold, $t_{th} \sim 22$ GeV for the 178.2 GeV beam, \cite{lynch2021experimental}. Given that we now have the threshold were the spectra due to hyperbolic recoil dominates over the synchrotron trajectory, let us now confirm the functional form of the acceleration produced by the recoil via conservation of momentum.

\section{Hyperbolic recoil acceleration from energy-momentum conservation} \label{econ}

Let us begin by examining the short time behavior of our hyperbolic recoil velocity, Eqn.~(\ref{rv}). Writing the initial velocity as, $v_{i} = v_{b}$, initial Lorentz gamma, $\gamma_{i}$, and $v_{f} = v$, we have,
\bqa
v &=&  \frac{v_{b}-\frac{at/\gamma}{\sqrt{1 + \lb \frac{at}{\gamma } \rb^{2}}}}{1-v_{b}\frac{at/\gamma}{\sqrt{1 + \lb \frac{at}{\gamma } \rb^{2}}}}   \non \\
v_{f} &\approx & v_{i} - \frac{a}{\gamma_{i}^3}t, \label{vel}
\eqa
where $a$ is the proper acceleration and $a/\gamma^3= a_{lab}$. In this manner, we have the standard velocity addition, $v_{f} = v_{i} -v_{a}$. By multiplying both sides of the equation by the total initial beam energy, $E_{i} = m\gamma_{i}$, we can convert this into a statement of conservation of momentum. Hence,
\bqa
\frac{E_{i}}{E_{f}}p_{f} &=&  p_{i} - \frac{ma}{\gamma_{i}^2}t \non \\
 p_{f} & =& p_{i} - \lbk p_{f} \left(  \frac{E_{i}}{E_{f}} - 1 \right) + \frac{ma}{\gamma_{i}^2}t   \rbk. 
\eqa
The quantity in the bracket can be identified with the momentum of the emitted photon, $k = \omega$. Solving for the rapidity, $\frac{at}{\gamma}$, of the recoil, we obtain the following,
\bqe
\frac{a_{i}t}{\gamma_{i}} = \frac{\gamma_{i}\omega}{m} - \frac{\gamma_{i}p_{f}}{m}\lbk  \frac{E_{i}}{E_{f}} - 1 \rbk\label{i}
\eqe
Here the acceleration/rapidity is denoted, $a_{i}$, on account of us multiplying by the initial beam energy and thus characterizes the rapidity as measured in the framed characterized by $\gamma_i$. Likewise, if we multiply Eqn. (\ref{vel}) by the final beam energy, $E_{f} = m \gamma_{f}$, we obtain the rapidity as measured in the frame characterized by $\gamma_f$,
\bqe
\frac{a_{f}t}{\gamma_{i}} = \frac{\gamma_{i}\omega}{m}\frac{E_{i}}{E_{f}} - \frac{\gamma_{i}p_{i}}{m}\lbk  \frac{E_{i}}{E_{f}} - 1 \rbk \label{f}
\eqe
We must note that both of the above expressions, Eqn. (\ref{i}) and Eqn. (\ref{f}), are identical provided that momentum is conserved, $p_{f} \approx p_{i} - \omega$. Then, by summing over both rapidities, $\frac{at}{\gamma_{i}} = \frac{a_{i}t}{\gamma_{i}}+\frac{a_{f}t}{\gamma_{i}}$, we obtain the rapidity as measured in the lab frame,
\bqe
\frac{at}{\gamma_{i}} = \frac{\gamma_{i}\omega}{m}\lbk \frac{E_{i}}{E_{f}}+1 \rbk- \frac{\gamma_{i}(p_{i}+p_{f})}{m}\lbk  \frac{E_{i}}{E_{f}} - 1 \rbk.
\eqe
Here we see that if we do not have conservation of energy, $E_{i} = E_{f}$, we then have the following rapidity,
\bqe
\frac{at}{\gamma_{i}}\Big|_{E_i = E_f} = \frac{2\gamma_{i}\omega}{m}. \label{rapne}
\eqe
In order to examine the rapidity with conservation of energy, we recall the relativistic momentum-energy relation, $p = \sqrt{E^2 -m^2}$, which has the following the expansion, $p \approx E - \frac{m^2}{2E}$, when $E \gg m$, i.e. the ultra relativistic regime. Applying this to for the initial and final momenta, we obtain
\bqe
\frac{at}{\gamma_{i}}\Big|_{E_i \neq E_f } = \frac{\omega}{2 }\frac{\lbk E_{i}+ E_{f}  \rbk}{E_{f}^2}. \label{rapwe}
\eqe
The time scale in the above rapidity is determined by the formation time since it characterizes the duration of motion and thus acceleration \cite{1986SvPhU..29..788A}. Due to the fact that there is recoil acceleration for all photon emission, we will consider the channeling formation time, $t_{f\Omega} = \frac{1}{\gamma \Omega}$, evaluated for a characteristic photon, $\omega = 2 \gamma^3 \Omega$. Thus we have $t_{f\Omega} = \frac{2 \gamma^2}{\omega}$ for $\omega \leq \omega_c$ and $t_{f\Omega} = \frac{2 \gamma^2}{\omega_c}$ for $\omega > \omega_c$. Let us first examine the case below threshold without energy conservation, Eqn. (\ref{rapne}),
\bqe
a = \frac{\omega^2}{m}.
\eqe
Here we have reproduced the standard recoil acceleration \cite{lynch2021experimental, 2024PhRvD.109j5009L, 2025GReGr..57..116L} which went into the asymptotic analysis. Next, let us examine the case with conservation of energy, Eqn. (\ref{rapwe}). Below threshold we have,
\bqe
a = \frac{\omega^2}{4 \gamma E_i} \frac{(2-\omega/E_i)}{(1-\omega/E_i)^2}.
\eqe
Note, here we made use of conservation of energy, $E_{f} = E_{i}-\omega$. Finally, by considering the case for photon frequencies above threshold, we then have,
\bqe
a = \frac{13 \omega }{144 \gamma}. \label{acc}
\eqe
Here we set our final energy, $E_{f} = E_i - \omega_c = \frac{6 E_{i}}{7}$. It is this hybrid recoil acceleration, Eqn. (\ref{acc}) which we shall use to analyze the experimental data on account of the fact that it is the initial acceleration experienced at threshold, see section (\ref{acce}) below for more details. Let us now develop the theoretical power spectrum which we will apply to the data sets.

\section{Ultra relativistic synchrotron emission: Classical, quasi-classical, and Unruh recoil}
The theory of ultra relativistic synchrotron emission is the standard formalism used to describe processes in strong field QED . Moreover, the quasi-classical formalism has served as a benchmarch for the incorporation of recoil, with classical synchrotron being reproduced in the ``no recoil" limit. We will follow the prescription of the quasi-classical formalism and incorporate the hyperbolic recoil into the trajectory. In this way, the no recoil limit will also yield our synchrotron theory with the Unruh effect. 

The probability for photon emission under ultra the relativistic synchrotron limit, which incorporates recoil via the quasi-classical formalism \cite{berestetskii1982quantum, 1998ephe.book.....B}, is given by 
\bqa
\mathcal{P} &=& \frac{\alpha}{4\pi^{2}} \int \frac{d^{3}k}{\omega}\int d t \int d t'  \non \\
& \times &\lbk \frac{E_{i}^2+E_{f}^2}{2E_{f}^2} \lbk \mathbf{v(t')}\cdot \mathbf{v(t)}-1\rbk+\frac{1}{2} \lb \frac{\omega m }{E_{f} E_{i}}\rb^2\rbk \non \\ 
& \times & e^{i \frac{E_{i}}{E_{f}}(\mathbf{k}\cdot \mathbf{\Delta x} -\omega \Delta t)}.
\eqa
Here, $\alpha  = \frac{q^2}{4 \pi \epsilon_{0}}$ is the fine structure constant. Thus, by examining specific trajectories, characterized by $\mathbf{v}(t)$ and $\Delta \mathbf{x} (t)$, we are able to compute the radiation spectrum both with and without quasi-classical recoil. The presence of the recoil correction is manifested by $E_{f} = E_{i} -\omega$. To recover the classical spectrum, we simply set $E_{f} = E_{i}$. By formulating the difference, $\xi = t'-t$, and average, $\eta = (t'+t)/2$, times we can then develop power spectrum, $\frac{d\mathcal{P}}{d \omega} = \frac{d P }{d\eta d\omega}\omega$. Hence,
\bqa
\frac{d\mathcal{P}}{d \omega} &=& \frac{\alpha}{4\pi^{2}} \int \omega^2 \sin{(\theta)}d\theta d\phi \int d \xi \non \\
 & \times & \Bigg[ \frac{E_{i}^2+E_{f}^2}{2E_{f}^2} \lbk \mathbf{v}(\eta + \xi/2)\cdot \mathbf{v}(\eta - \xi/2)-1\rbk\non \\
 &+&\frac{1}{2}\lb \frac{\omega m }{E_{f} E_{i}}\rb^2 \Bigg]\non \\
 & \times & e^{i (\mathbf{k}\cdot \mathbf{\Delta x} -\omega \xi)}. \label{rate}
\eqa
We will now compute the power spectrum for the combined synchrotron and hyperbolic/Unruh trajectory,
\bqa
v_{z}(t) &=&  \frac{v_{0}\cos{(\Omega t)}-\frac{at/\gamma}{\sqrt{1 + \lb \frac{at}{\gamma } \rb^{2}}}}{1-v_{0}\cos{(\Omega t)}\frac{at/\gamma}{\sqrt{1 + \lb \frac{at}{\gamma } \rb^{2}}}}\non \\
v_{\perp}(t) &=& -v_{0}\sin{\Omega t}.
\eqa
By noting our original time coordinates are written as, $t' = \eta +\xi/2$ and $t = \eta -\xi/2$, we will then have the following expansion,
\bqa
\mathbf{v(\eta + \xi/2)}\cdot \mathbf{v(\eta - \xi/2)} \approx \hspace{4cm} \non \\
  1- \lb \frac{1}{\gamma} \rb^{2} - \frac{1}{2} \lbk \Omega^2 + \frac{a^2 }{\gamma^4 } \rbk \xi^2. \hspace{.5cm} \label{dv}
\eqa
We will also have our expansion in the trajectory, $\Delta \mathbf{x}$. Here, we will only need to consider the z component. Hence,
\bqa
\Delta z & \approx & v_{0} \xi - \frac{a}{4 \gamma^3 } \xi^2 - \lbk \frac{a^2}{12\gamma^4 } + \frac{1}{24}\Omega^2   \rbk \xi^{3} \non \\
& \approx & \tilde{v} \xi \hat{z} - \lbk \frac{a^2}{12\gamma^4 } + \frac{1}{24}\Omega^2   \rbk \xi^{3} \hat{z}. \label{dz}
\eqa
In the last line we made use of the fact that, $v_{f}(\xi)= v_{0} - \frac{a}{2\gamma^3 }\xi$, to combine the quadratic acceleration component into the linear component to formulate the average velocity, $\tilde{v} = (v_{0}+v_{f})/2 $. By defining $\omega'  =  \frac{E_{i}}{E_{f}} \omega$, $a_{0} = \lb \frac{a^2\omega'}{4\gamma^4 } + \frac{\omega'}{8}\Omega^2   \rb^{1/3}$, $z = \frac{\omega'}{a_{0}}(1-\tilde{v}) = \frac{\omega'}{2a_{0}\gamma^2}$, and making use of the standard Airy function identities \cite{berestetskii1982quantum}, we arrive at our power spectrum, 
\bqa
\frac{dP}{d\omega}  &=& -\frac{\alpha \omega }{a_{0}}   \Bigg[ \lbk \frac{E_{i}^2+E_{f}^2}{4E_{f} E_{i} }\rbk \lbk \Omega^2 + \frac{a^2 }{\gamma^4 } \rbk \frac{1}{a_{0}}Ai'(z) \non \\
&+ & \frac{a_{0}}{\gamma^2 }\int_{z}^{\infty}Ai(z')dz' \Bigg]. \label{spec}
\eqa
It is this spectrum, with its various limiting cases to the classical, quasi-classical, and Unruh/recoil theory, that we shall now apply to the CERN-NA63 data sets.

\subsection{Parameters for experimental analysis}
In order to examine the experimental data of CERN-NA63 we note that there are only two free parameters which govern the hyperbolic/Unruh spectra; the acceleration, $a$, and the channeling frequency, $\Omega$. These two parameters are completely fixed by the threshold frequency, $ \omega_c = E_{i}/7$. Thus we will have our acceleration from Eqn. (\ref{acc}), and the channeling frequency, $\Omega = \frac{\omega_c}{2\gamma^3}$. Hence,
\bqa
a &=& \frac{13 \omega }{144 \gamma} \non \\
\Omega &=& \frac{E_i}{14\gamma^3}.
\eqa
We will also fix the overall scale by normalizing the Unruh spectra and the data over the interval: $\omega_c \rightarrow \omega_{max}$. The chi-squared per degree of freedom we present are computed via $\chi^{2}/\nu = \sum_{\omega_c}^{\omega_{max}} \frac{\lb y_{theory} - y_{exp} \rb^2}{\Delta y_{exp} (\nu-2)}$. Here, $\Delta y_{exp}$ is the error and $\nu = \omega_{max} - \omega_c + 1$ is the number of degrees of freedom or data points. Note, we have included the normalization constant for the theory as an additional parameter for the chi-squared statistic and have presented $\chi_{U}^{2}/\nu $, $\chi_{C}^{2}/\nu $, and $\chi_{QC}^{2}/\nu $ for the Unruh, classical, and quasi-classical theories respectively. Finally, we note that with the above acceleration and channeling frequency, all spectra are completely fixed and, beyond the normalization constant, are parameter free. Let us now turn to the development of the power spectra which we will compare to the CERN-NA63 data sets. 

\subsection{Comparison of synchrotron spectra}
In this section, we will examine the three spectra of synchrotron emission, Eqn. (\ref{spec}): classical, quasi-classical, and Unruh recoil. To obtain the classical spectrum, we set $E_{f} = E_{i}$ along $a = 0$, the quasi-classical spectrum will have $a = 0$ and $E_{f} = E_{i} - \omega$, and the Unruh recoil spectrum will have $E_{f} = E_{i}$, $a =\frac{13 \omega }{144 \gamma}$. For all spectra, we have $\Omega = \frac{E_i}{14\gamma^3}$. The overall qualitative features are contained in Fig. \ref{comp1} below. For different beam energies, the only effect is to shift the overall scale of the normalized power spectrum up or down. The shape of all spectra is completely fixed by $\omega_c$. 

\begin{figure}[H]
\centering  
\includegraphics[scale=.26]{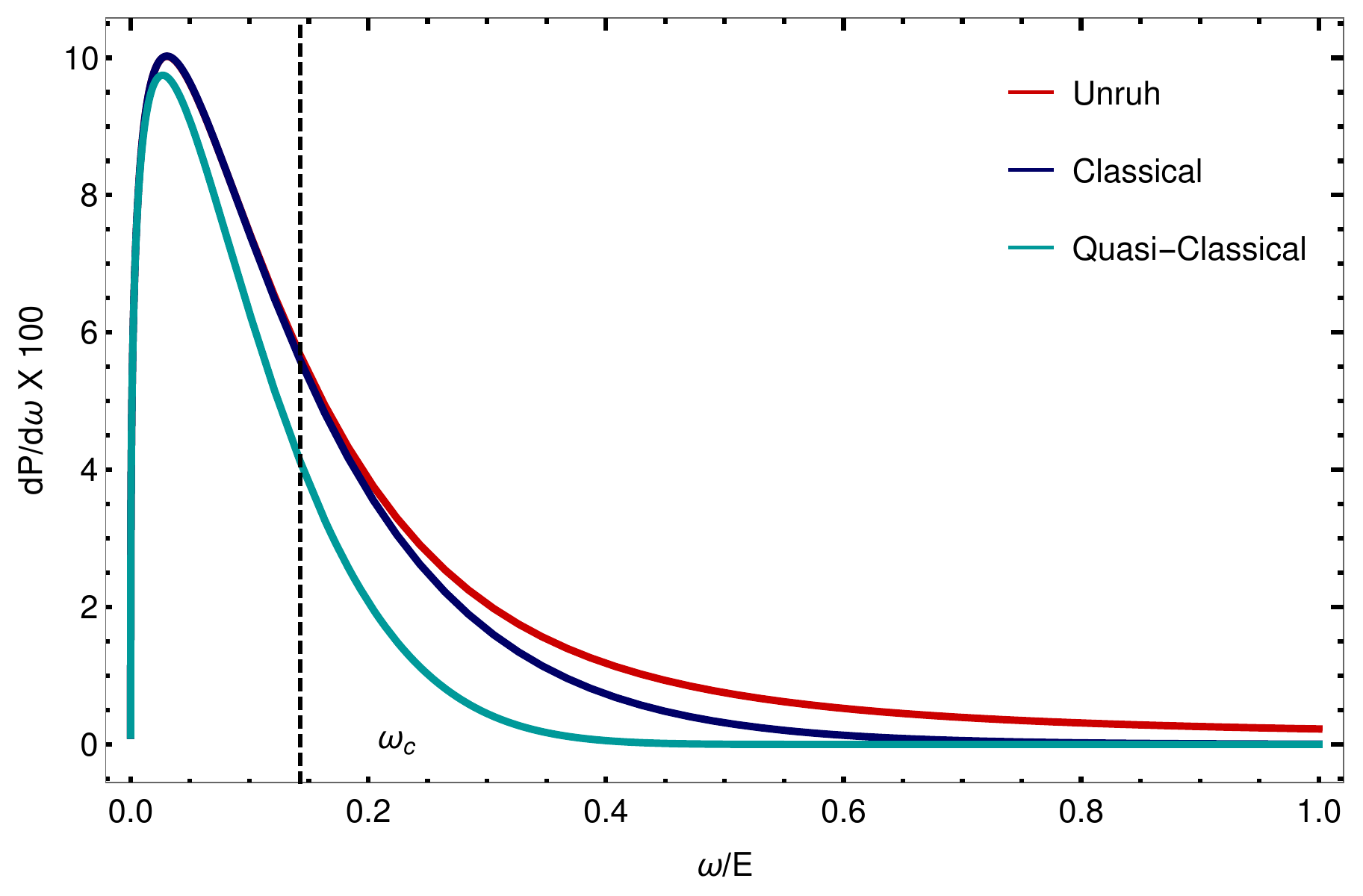}
\caption{The normalized power spectrum for all three models with a beam energy, $E=100$ GeV.} 	
\label{comp1}
\end{figure}

Note, the effect of the quasi-classical recoil to steepen the high energy tail of the spectrum and the effect of the Unruh recoil is to enhance the high energy tail. Let us now turn to the NA63 experimental data and apply the above spectra to their data sets. In Figures (\ref{38spec}-\ref{62spec}) below, we present two 178.2 GeV (3.8 mm and 10 mm) and five 50 GeV (1.1 mm, 2 mm, 4.2 mm, and two 6.2 mm) channeling radiation data sets \cite{2018NatCo...9..795W, 2019PhRvR...1c3014W}. The difference between the two 6.2 mm data sets is a different incident positron beam angle on the crystal sample \cite{2019PhRvR...1c3014W}. What is of particular note is the enhancement in the high frequency tail of all data sets which departs from the classical synchrotron theory above the critical frequency where the hyperbolic recoil, i.e. the Unruh effect, takes hold. Moreover, all parameters of the analysis are completely fixed by the critical frequency and we thereby have a relatively simple technique to determine these characteristic parameters based simply on the beam energy. To characterize how well each theory matches the data sets we have averaged over all chi-squared statistics to present an overall figure of merit for each theory: $\chi_{U}^2/\nu = 2.12$, $\chi_{C}^2/\nu = 8.09$, and $\chi_{QC}^2/\nu = 47.0$.

\begin{figure}[H]
\centering  
\includegraphics[scale=.27]{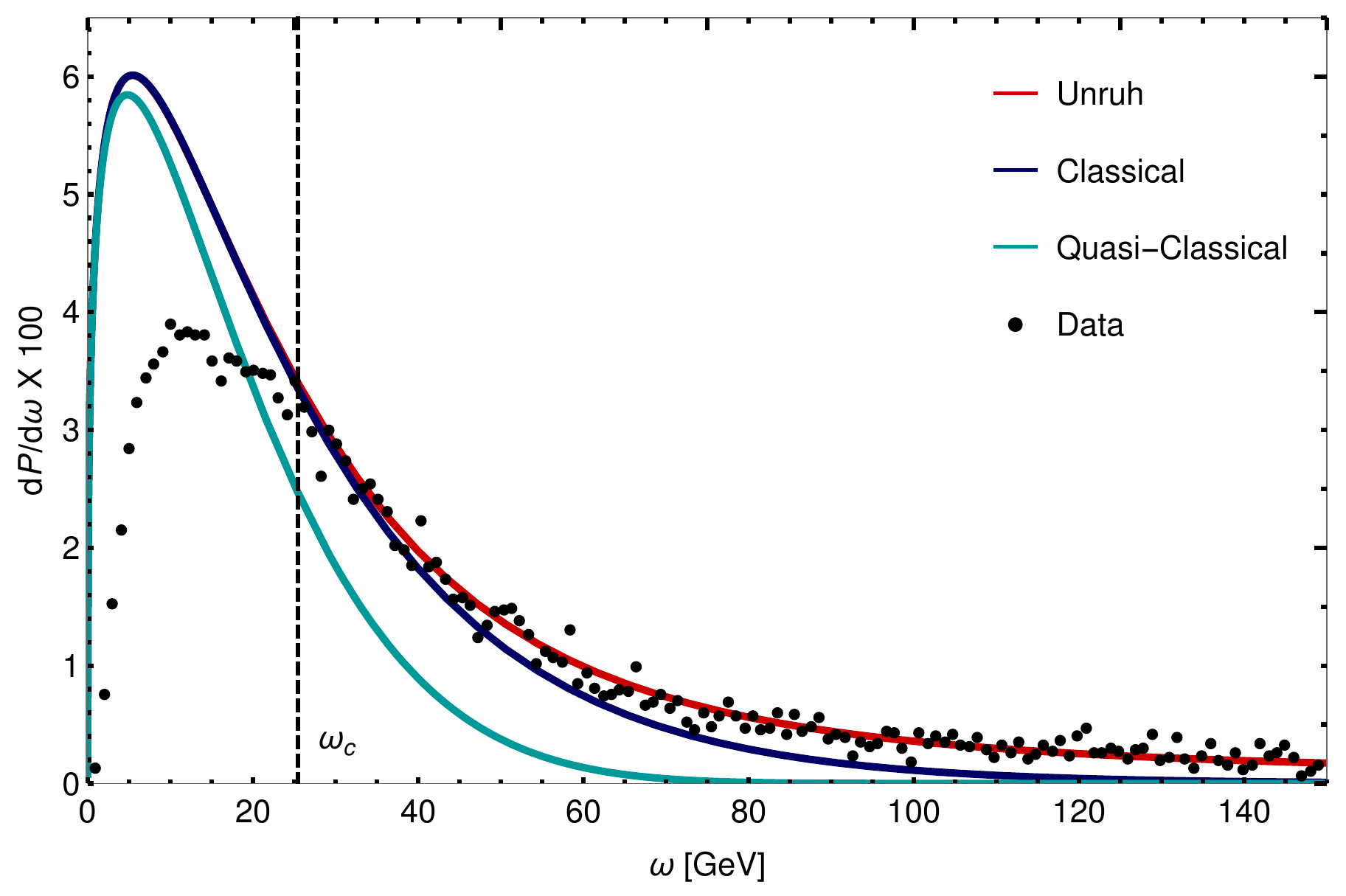}
\caption{The normalized power spectrum for the 178.2 GeV, 3.8 mm crystal with a $\chi_{U}^2/\nu = 1.38$, $\chi_{C}^2/\nu = 6.08$, and $\chi_{QC}^2/\nu = 34.3$.}
\label{38spec}
\end{figure}

\begin{figure}[H]
\centering  
\includegraphics[scale=.27]{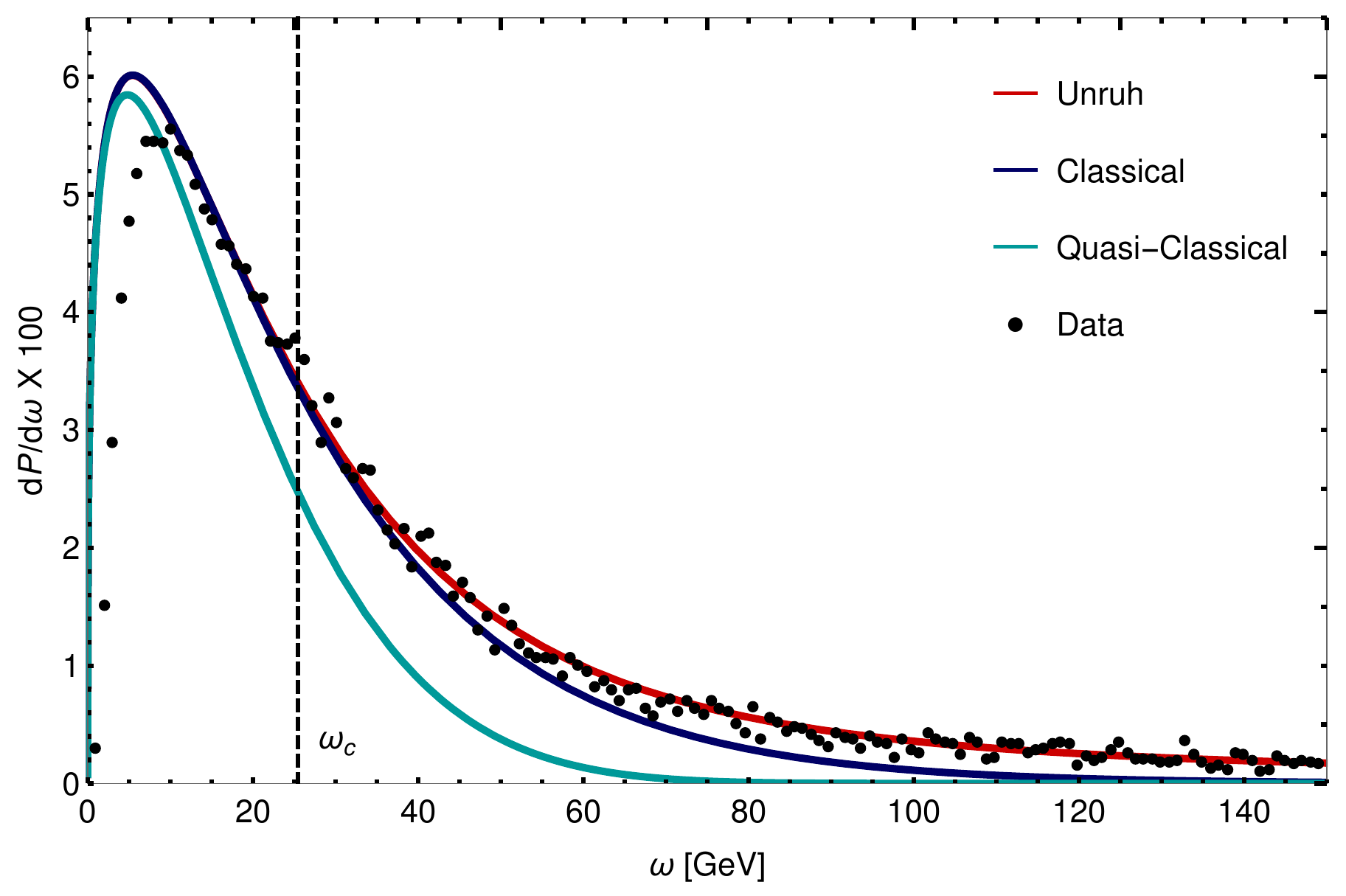}
\caption{The normalized power spectrum for the 178.2 GeV, 10 mm crystal with a $\chi_{U}^2/\nu = 2.11$, $\chi_{C}^2/\nu = 11.3$, and $\chi_{QC}^2/\nu = 72.2$.} 	
\label{10spec}
\end{figure}

\begin{figure}[H]
\centering  
\includegraphics[scale=.27]{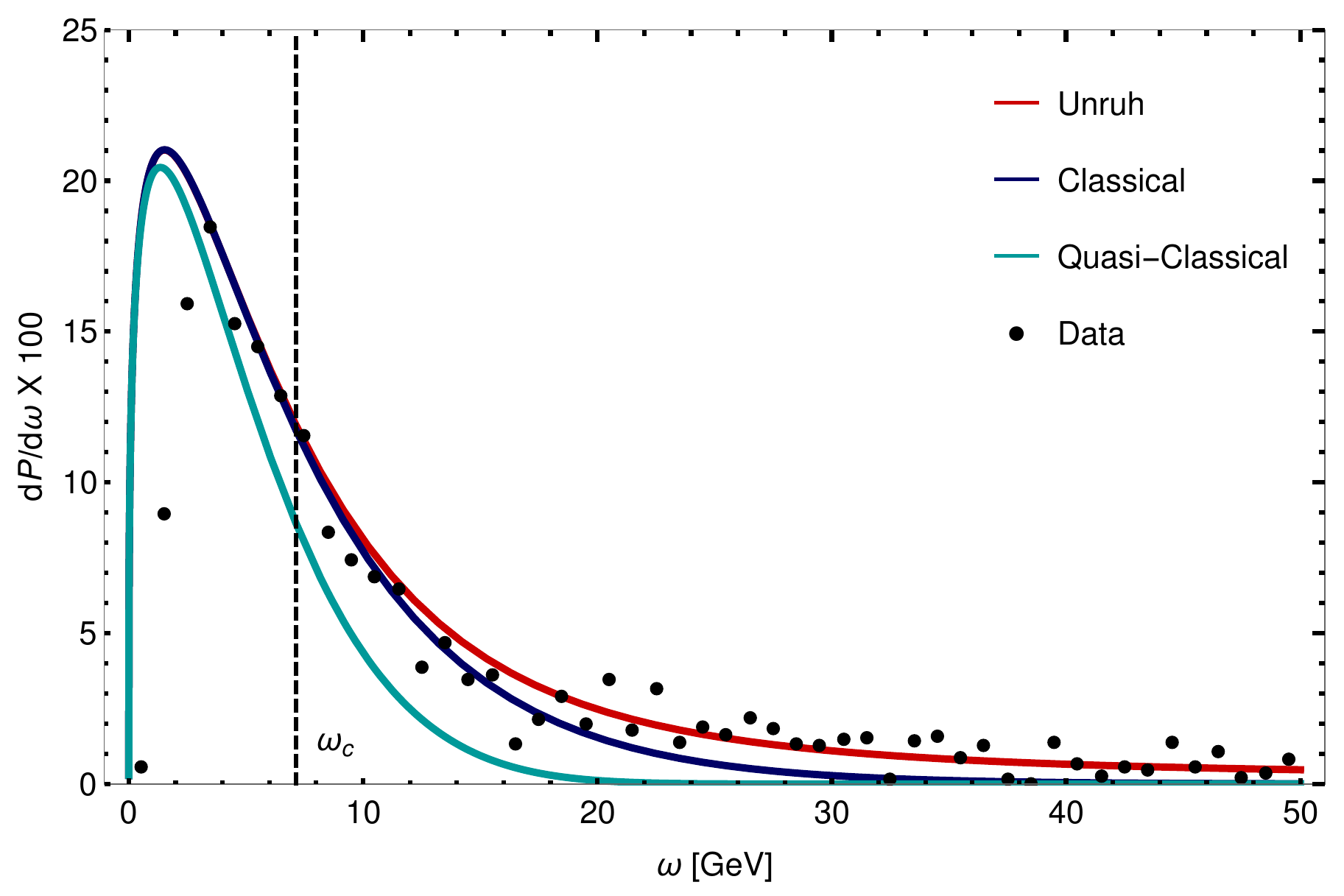}
\caption{The normalized power spectrum for the 50 GeV, 1.1 mm crystal with a $\chi_{U}^2/\nu = 3.07$, $\chi_{C}^2/\nu = 5.20$, and $\chi_{QC}^2/\nu = 20.5$.} 	
\label{1spec}
\end{figure}

\begin{figure}[H]
\centering  
\includegraphics[scale=.27]{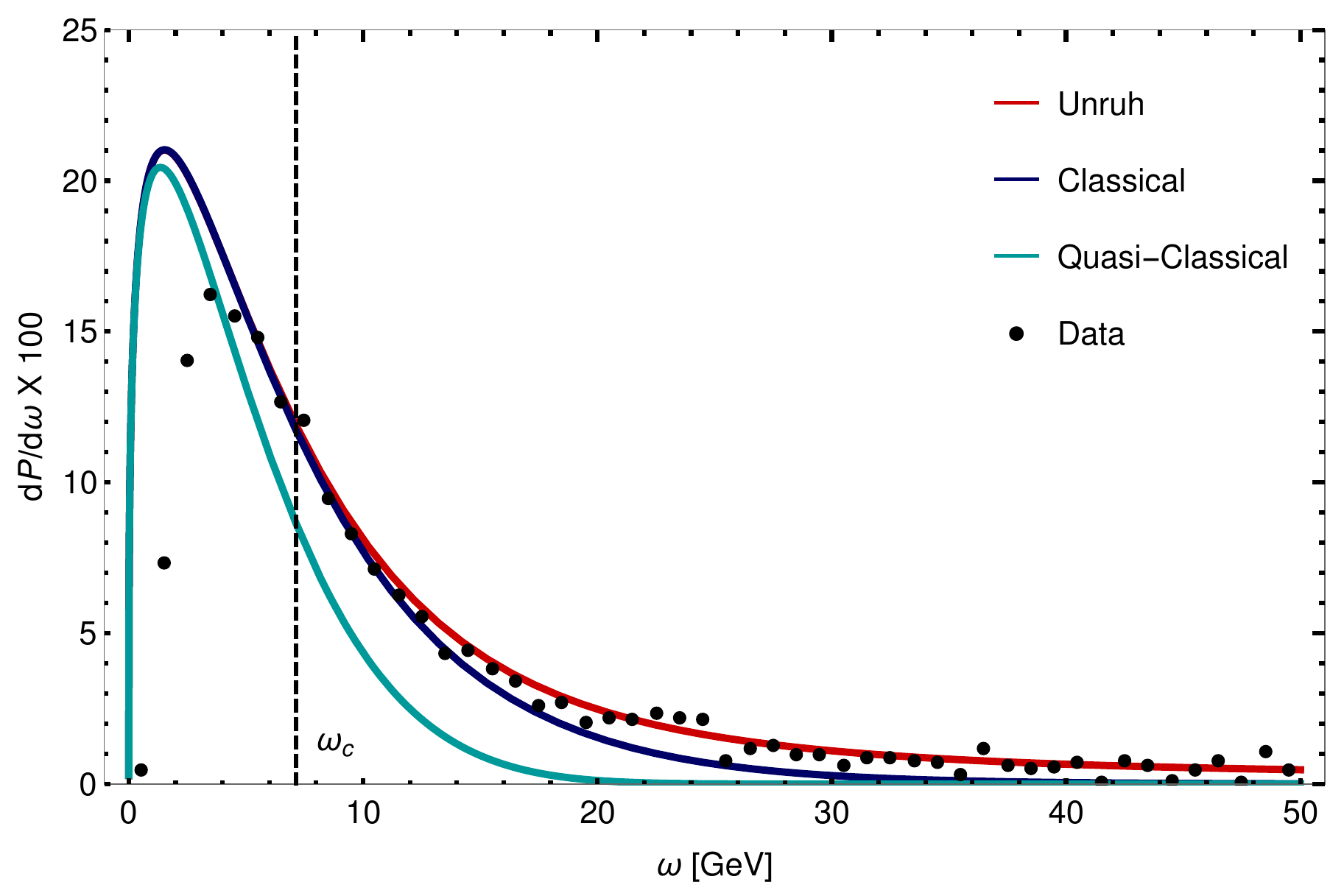}
\caption{The normalized power spectrum for the 50 GeV, 2 mm crystal a $\chi_{U}^2/\nu = 2.79$, $\chi_{C}^2/\nu = 11.5$, and $\chi_{QC}^2/\nu = 73.6$.} 	
\label{2spec}
\end{figure}

\begin{figure}[H]
\centering  
\includegraphics[scale=.27]{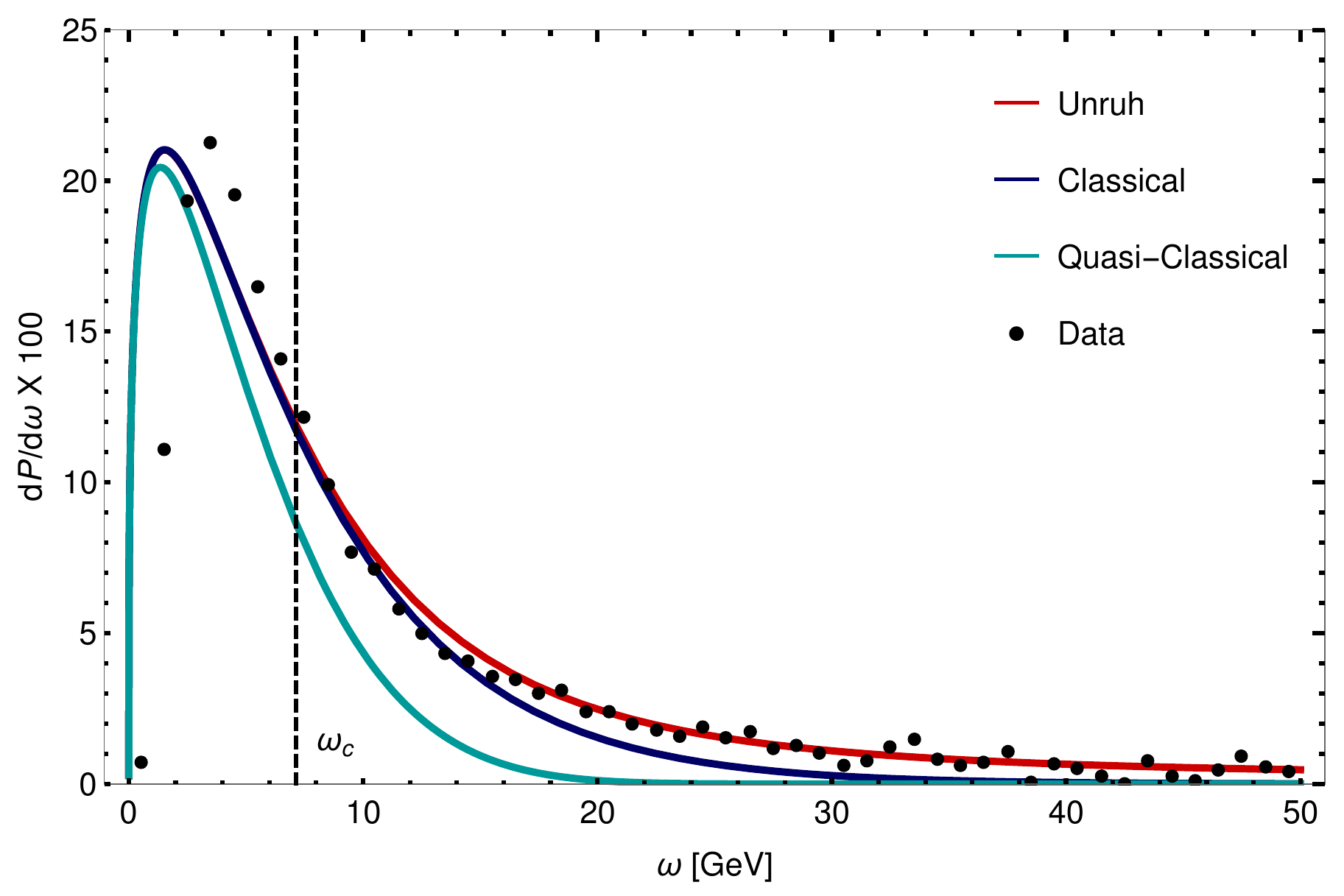}
\caption{The normalized power spectrum for the 50 GeV, 4.2 mm crystal with a $\chi_{U}^2/\nu = 2.11$, $\chi_{C}^2/\nu = 7.60$, and $\chi_{QC}^2/\nu = 44.5$.} 	
\label{4spec}
\end{figure}

\begin{figure}[]
\centering  
\includegraphics[scale=.27]{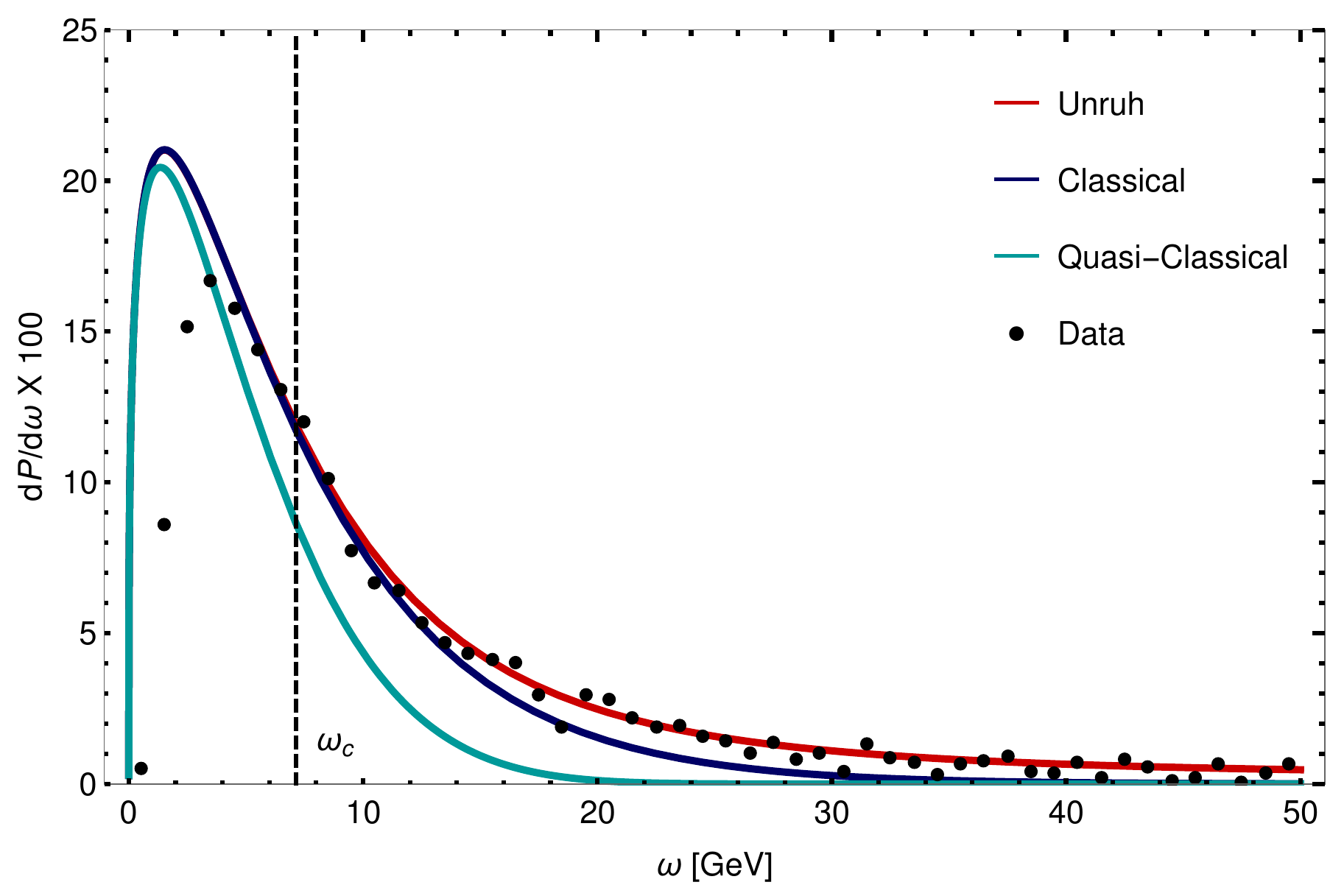}
\caption{The normalized power spectrum for the 50 GeV, 6.2-1 mm crystal with a $\chi_{U}^2/\nu = 1.78$, $\chi_{C}^2/\nu = 6.17$, and $\chi_{QC}^2/\nu = 36.7$.} 	
\label{61spec}
\end{figure}

\begin{figure}[]
\centering  
\includegraphics[scale=.27]{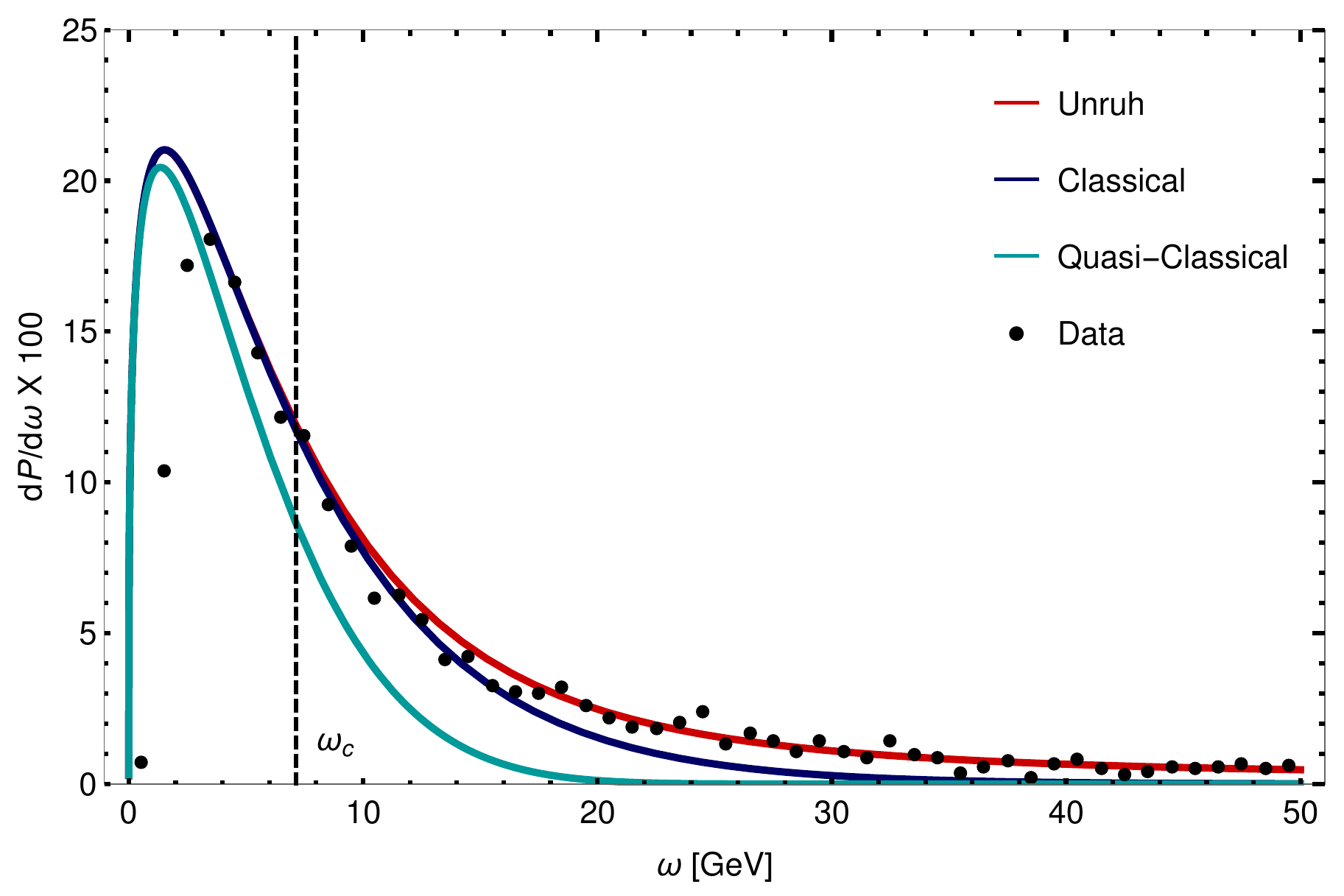}
\caption{The total power spectrum for the 50 GeV, 6.2-2 mm crystal with a $\chi_{U}^2/\nu = 1.57$, $\chi_{C}^2/\nu = 8.78$, and $\chi_{QC}^2/\nu = 47.0$.} 				
\label{62spec}
\end{figure}

\section{Discussion}
The above power spectra comparisons provide evidence for the presence of the Unruh effect in the high frequency tail. In the analysis, we utilized a 1-dimensional recoil acceleration. The validity of this is examined in more detail below. We also examine how the Unruh effect manifests itself in the standard theories of radiation reaction and guiding center recoil. Finally, we will examine how frequency dependent temperatures presents themselves in the experimental data. 

\subsection{1-dimensional recoil} \label{1d}
One of the key assumptions in the analysis is that the recoil is effectively 1 dimensional, i.e. along the beam axis. In this regard, let us examine this in more detail for both the laboratory and beam frames. Recalling the recoil velocity is given by $v_{a} =  \frac{at/\gamma}{\sqrt{1 + \lb \frac{at}{\gamma } \rb^{2}}}$, we then have components along the beam direction, $v_{az} = v_{a} \cos{(\theta')}$, and the perpendicular axis, $v_{a\perp} = v_{a} \sin{(\theta')}$. Here, the angle $\theta'$ is defined in the beam frame relative to the z-axis. In the lab frame, we have a final velocity along the z-axis, given by $v_{fz} = \frac{v_z+v_{az}}{1+v_z v_{az}}$, and a final velocity along the perpendicular axis given by $v_{f\perp} = \frac{v_{a\perp}/\gamma}{1+v_z v_{az}}$. Note, here we have kept the direction of the recoil arbitrary and determined by the angle $\theta '$. 

\begin{figure}[]
\centering  
\includegraphics[scale=.28]{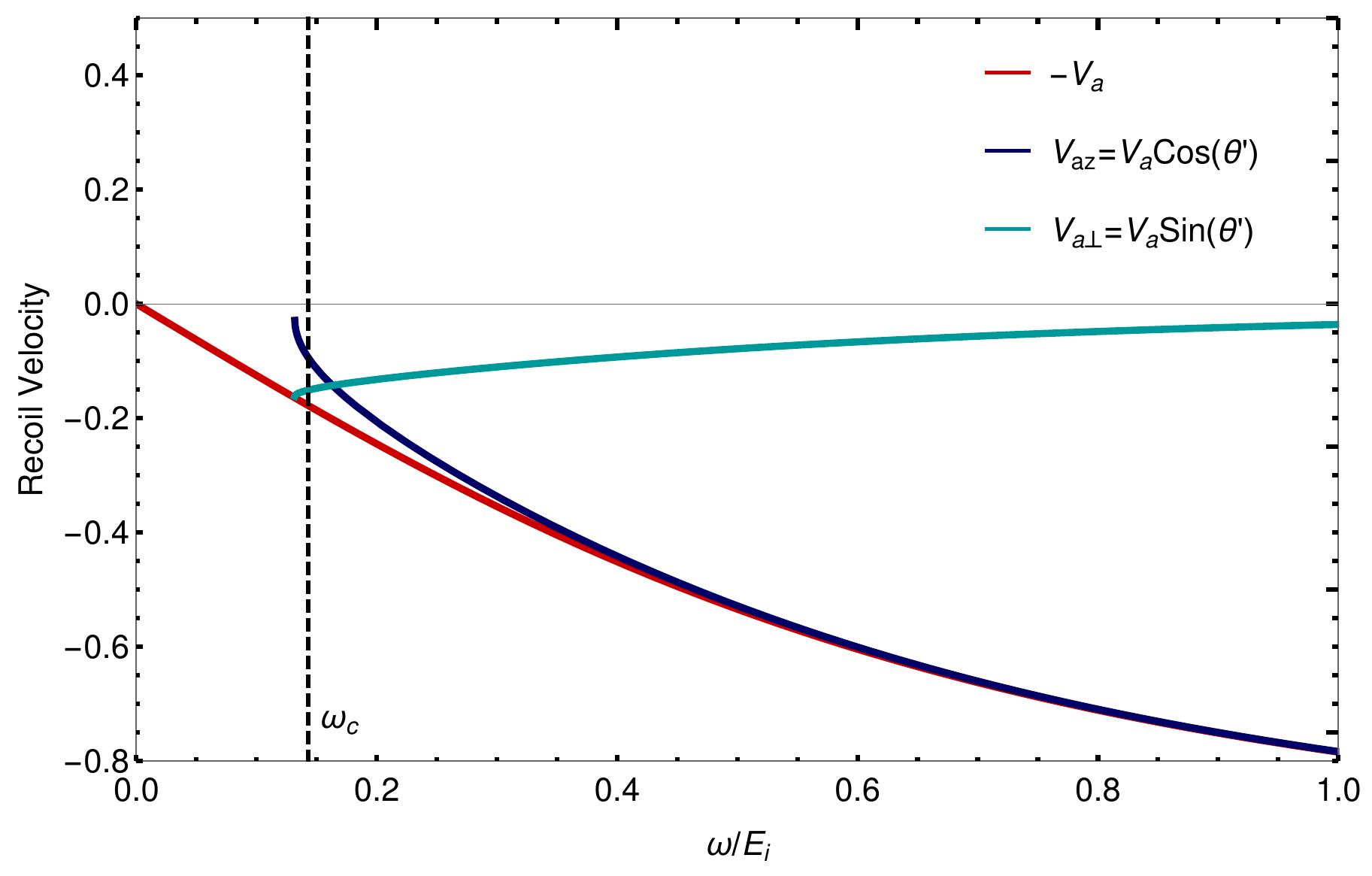}
\caption{The various components of the beam frame recoil velocity as a function of photon energy $\omega'$. As $\omega': 1/7 \rightarrow 1$, the perpendicular component $v_{a\perp} \rightarrow 0$ and the z component converges to $v_{az} \rightarrow -v_{a}$.}  			
\label{velocity}
\end{figure}

The angle of the final positron velocity can be determined via, $\frac{v_{f\perp}}{v_{fz}} = \frac{v_{a\perp}/\gamma}{v_z+v_{az}} = \tan{(\theta)} $. Here, $\theta$, is defined in the lab frame with respect to the z-axis. In terms of the final momenta, we have, $p_{fz} = p_{i} - \omega \cos{(\theta)}$ and $p_{f\perp} = - \omega \sin{(\theta)}$. Since the final momenta also determine the same direction as the velocity, i.e. $\frac{v_{f\perp}}{v_{fz}} = \frac{p_{f\perp}}{p_{fz}}$, we have the following relationship between the photon energy and beam frame emission angle;
\bqe
\sin{(\theta')} + \bar{\omega} \cos{(\theta ')} = - \frac{\bar{\omega}}{v_{a} }.
\eqe
Note, here we have set $v_{i} = 1$, $\sin{(\theta)} =\frac{1}{\gamma}$, and $\cos{(\theta)} = \beta$ since the positron is ultra relativistic and defined $ \bar{\omega} = \frac{\omega}{E_{i} - \omega}$. Note, here we have selected that the photon will be emitted above the z-axis since $\sin(\theta)>0$. Next, we recall that we have our Unruh threshold at $\omega_c = \frac{E_{i}}{7}$. As such, we have $\bar{\omega} = \frac{1}{6}$. Likewise, utilizing our acceleration, Eqn. ({\ref{acc}}), and the time scale $t = \frac{2\gamma^2}{\omega_c}$, we have a velocity as a function of photon frequency given by,
\bqe
v_{a} = \frac{\frac{91 \omega'}{72}}{\sqrt{1 + \lb \frac{91 \omega'}{72}\rb^2}}.
\eqe
Here we have defined the dimensionless frequency, $\omega' = \frac{\omega}{E_{i}}$. With these parameterizations, we can determine the beam frame emission angle, as a function of recoil velocity, to be,
\bqe
\theta_{\pm }' = -i \ln{\frac{(1+6 i)v_a \pm (6-i)\sqrt{v_{a}^2 (37 v_{a}^{2}-1)}}{37 v_{a}^2}}.
\eqe
Here the $+$ sign corresponds to the case of recoil, with a velocity in the negative z direction, i.e. $\theta_+ : -\pi/2 \rightarrow -\pi$. As we shall see, there is a rapid convergence to $-\pi$. The forward scattering branch, $\theta_-$, is an unphysical solution corresponding to a positron which increases energy upon radiating, see Fig. \ref{angle} below. The fact we have a negative recoil angle, $\theta_{+}' <0$ is also consistent with the fact that the lab frame photon emission angle is positive, $\theta >0$. Dropping the $+$ subscript and simply using, $\theta_{+}' = \theta'$, we can also examine the various components of the recoil velocity, $v_{az} = v_{a}\cos{(\theta')}$ and $v_{a\perp} = v_{a}\sin{(\theta')}$, see Fig. \ref{velocity}. We see the perpendicular component converges to zero while the z component converges to $v_a$. As such, we have in the beam frame $v_{az} \sim v_{a}$ with an error of about $\frac{v_{az}}{v_a} \sim .5$ at threshold, $\omega' = \frac{1}{7}$, but rapidly converges to $\frac{v_{az}}{v_a} > .9$ for $\omega' = \frac{1}{7}+.1$ and beyond. Thus, in the beam frame, we have a rapid convergence to a 1 dimensional recoil along the z-axis.

In the lab frame, the situation is much simpler. The transverse component $v_{f\perp} = \frac{v_{a\perp}/\gamma}{1+v_z v_{az}}$ is already suppressed by a factor of $\frac{1}{\gamma}$ and is thus negligible. The z component, $v_{fz} = \frac{v_z+v_{az}}{1+v_z v_{az}}$, depends on the z component in beam frame, $v_{az}$, which we know rapidly converges to $v_{az} \cos{\theta '} \rightarrow -v_{a}$. Here the minus sign is due to the angle converging to $-\pi$. Thus, we have $v_{fz} = \frac{v_z-v_{a}}{1-v_z v_{a}}$, which is the 1 dimensional lab frame recoil velocity used in the analysis. We must also comment on the fact that kinematically, the allowed solutions for the beam frame emission angle start out at $\omega' = \frac{12}{92} \sim .13$. This also reflects the fact that the asymptotic analysis yields an Unruh threshold of $\omega' = \frac{1}{7} \sim .14$. Thus, there is consistency between the kinematic and time scale analyses.

\begin{figure}[]
\centering  
\includegraphics[scale=.28]{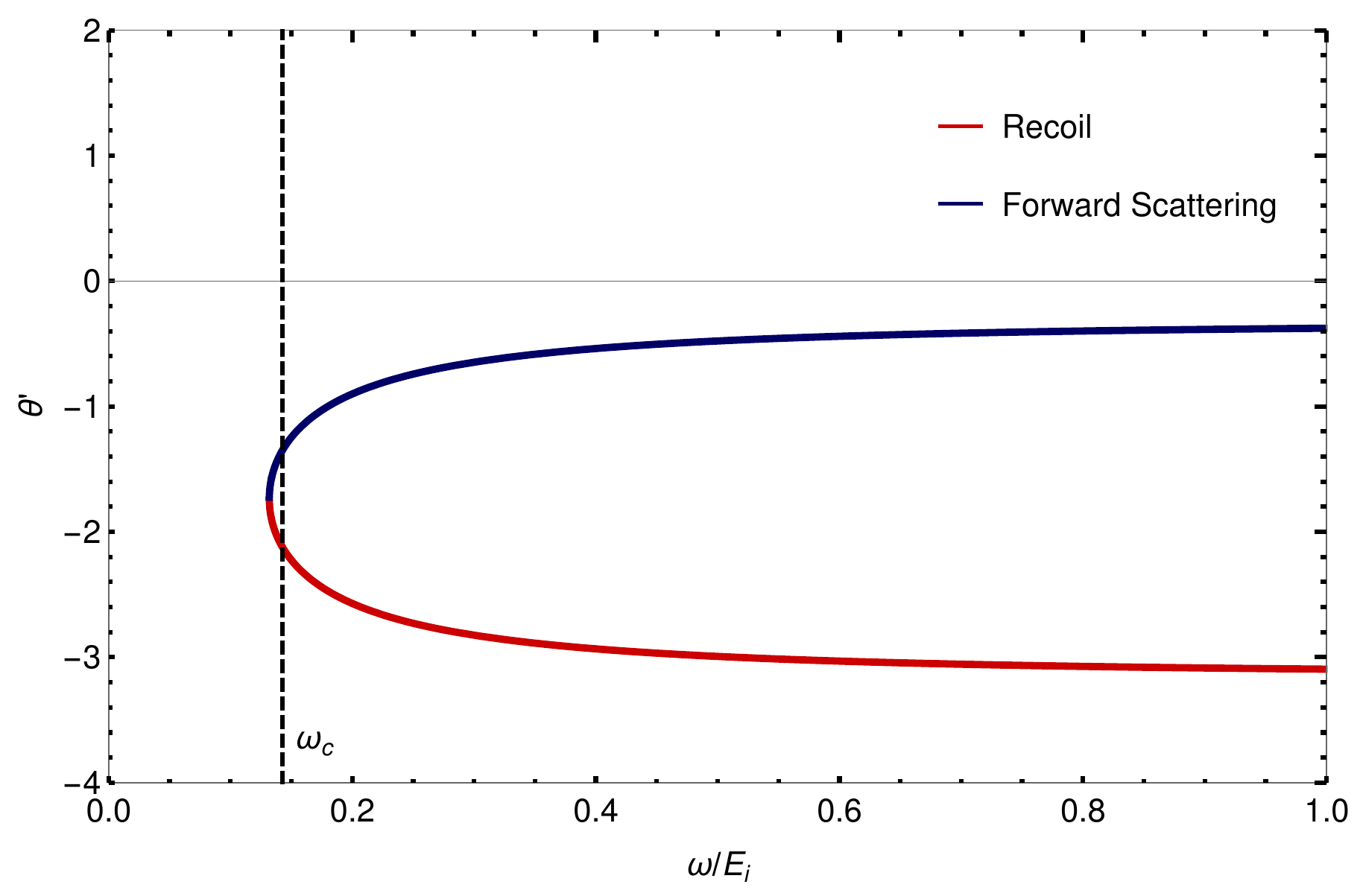}
\caption{The allowed beam frame emission angles based on the kinematics of relativistic velocity addition. The lower branch, in red, corresponds to a positron velocity due to the recoil from photon emission with energy $\omega'$. We see the emission angle starts at $-\pi/2$ and rapidly converges to $-\pi$. The upper branch, in blue, corresponds to the unphysical velocity increase upon photon emission. }  			
\label{angle}
\end{figure}

\subsection{Guiding center recoil: the Unruh effect from the Landau-Lifshitz equation} \label{ll}
One of the key insights from the observation of radiation reaction comes from its relation to the Unruh effect. The first experimental measurements of radiation reaction consisted of two experiments utilizing ultra intense lasers \cite{2018PhRvX...8a1020C, 2018PhRvX...8c1004P} and the NA63 channeling experiments \cite{2018NatCo...9..795W}. The key insight derived by these measurements was an overall indication of either a quantum radiation reaction, i.e. motion determined by the Lorentz force but with a quantum mechanical recoil subtracted, or dynamics governed by the Landau-Lifshitz equation \cite{1975ctf..book.....L} of motion. Moreover, follow up experiments by NA63 even favored the experimental verification of the Landau-Lifshitz (LL) equation \cite{2021NJPh...23h5001N}. Thus, in systems with radiation reaction described by the LL equation, we now have an experimental insight into the relationship between radiation reaction and the Unruh effect.

During the 1980s, a series of papers was devoted to investigating a discrepancy that was found in both the conservation of energy and angular momentum in synchrotron emission. \cite{1983JPhA...16L.669L, 1984JPhA...17L..91L, 1984JPhA...17L.223L, 1984JPhA...17.2895D, 1985JPhA...18L.111W, 1987JPhA...20.2105P, 1987JPhA...20.2405L, 1987A&A...176L..21L}. The resolution of this apparent paradox came in the form of including guiding center recoil into the analysis of synchrotron emission. Broadly speaking, when the radiation produced by a positron undergoing synchrotron motion becomes intense, i.e. photons on the order of the beam energy, the circular arc swept out by they cyclotron motion can shift in response to the photon recoil. This is in addition to the standard radiation reaction which leads to inspiral as well as an undulation produced by the $\sim \pm \frac{1}{\gamma}$ emission angle. It was found that the incorporation of guiding center recoil resolves the apparent paradoxes of energy-momentum conservation \cite{1985JPhA...18L.111W, 1987JPhA...20.2105P} 

The inclusion of this guiding center recoil into the theory of synchrotron radiation emission was developed by considering a charged particles motion under the LL equation of motion \cite{1987JPhA...20.2405L, 1987A&A...176L..21L}. The resultant trajectory, which includes recoil, is then used as the vector current source for radiation emission. Let us consider the fully relativistic Lorentz-Abraham-Dirac (LAD) equation of motion,
\bqe
m\frac{d u^{\m}}{d\tau} = q F^{\m \nu}u_{\nu} +\frac{2}{3}q^2\lbk \frac{d^2 u^{\m}}{d\tau^2}  +\frac{d u^{\nu}}{d\tau}\frac{d u_{\nu} u^{\m}}{d\tau} \rbk .
\eqe 
Here the positron 4-velocity is given by $u^{\m}$, the standard Lorentz force is given by $q F^{\m \nu}u_{\nu}$ and the radiation reaction force is the term in brackets. Then, by substitution of the zeroeth order four acceleration, $\frac{d u^{\m}}{d\tau} = \frac{q}{m} F^{\m \nu}u_{\nu}$, into the radiation reaction force, we arrive at the LL equation,
\bqa
m\frac{d u^{\m}}{d\tau} &=& q F^{\m \nu}u_{\nu} + \frac{2}{3}q^2 \Bigg[ \frac{q}{m} \lb \partial_{\alpha} F^{\m \nu}\rb u^{\alpha} u_{\nu} - \frac{q^2}{m^2}  F^{\m \nu} F_{\alpha \nu} u^{\alpha} \non \\
&+& \frac{q^2}{m^2} \lb  F^{\alpha \nu} u_{\nu} \rb \lb  F_{\alpha \lambda} u^{\lambda} \rb u^{\m} \Bigg] .
\eqa
Although the above equation appears somewhat cumbersome, there are analytic solutions for special cases, e.g. for a charged particle in a laser field \cite{2008LMaPh..83..305P, 2010PhRvwisD..82i6012H}. For the case of a homogeneous magnetic field in the positive z direction, $\mathbf{B} = B \hat{z}$, the LL equation reduces to the two coupled equations \cite{1987JPhA...20.2405L, 1987A&A...176L..21L},
\bqa
\frac{d p_{x}}{dt} &=& -m(\Omega_c v_{y}+\bar{a} \gamma^2 \beta_x) \non \\
\frac{d p_{y}}{dt} &=& m(\Omega_c v_{x}-\bar{a} \gamma^2 \beta_y).
\eqa
Here $\Omega_c = \frac{q B}{m} $ is the cyclotron frequency and the parameter $\bar{a} = \frac{2 \alpha \hbar \Omega_{c}^2}{3 m c}$ characterizes the recoil acceleration from the radiation reaction with $\alpha$ being the fine structure constant. The above coupled equations can be solved to yield the synchrotron trajectory with the recoil correction,
\bqa
\beta_{z}(t) &=& \lbk 1-\frac{1}{2} \lb \frac{\bar{a} t}{c}+\frac{1}{\gamma_{i} }  \rb^{2}  \rbk \cos{\lbk \frac{\bar{a} \Omega_{c} t^2}{2c} + \frac{\Omega_c t}{\gamma_{i}}\rbk }   \non \\
\beta_{\perp}(t) &=& \lbk 1-\frac{1}{2} \lb\frac{\bar{a} t}{c}+\frac{1}{\gamma_{i} }  \rb^{2}  \rbk \sin{\lbk \frac{\bar{a} \Omega_{c} t^2}{2c} + \frac{\Omega_c t}{\gamma_{i}}\rbk } .
\eqa
Note, in the above expressions we changed our axes $x \rightarrow z$ and $y \rightarrow \perp$ to directly compare to the synchrotron trajectory with hyperbolic recoil that was used in our analysis. In this regard, let us consider the short time behavior of the above trajectory,
\bqa
v_{z}(t) & \approx & v_{i} - \frac{\bar{a} }{\gamma_{i}} t \non \\
v_{\perp}(t) &\approx & 0.
\eqa
From the above expansion, we can immediately ascertain that the proper acceleration, by comparison with Eqn. (\ref{vel}), produced by the recoil is given by $\bar{a} = a/ \gamma_{i}^2$. Moreover, we can define the fractional energy loss of the positron by $\m = \gamma_{i} \frac{\bar{a}}{\Omega_{c}}$ \cite{1987JPhA...20.2405L, 1987A&A...176L..21L}, and note that it is an invariant parameter which characterizes the field strength of the system, $\m 	= \frac{2}{3} \alpha \chi$. We can then determine the proper acceleration to be, $a = \m \gamma_{i}\Omega_{c} $. Finally, recalling that the lab frame orbital frequency, $\Omega_{0}$, in the relativistic limit is related to the cyclotron frequency by $\Omega_{0} = \Omega_{c}/\gamma_{i}$ and this quantity is equivalent to the lab frame channeling oscillation frequency $\Omega_0 = \Omega$. Thus, mapping the lab frame cyclotron frequency to the emitted photon frequency, $\omega = 2 \gamma_{i}^3 \Omega_{0}$ yields, $\omega = 2 \gamma^2 \Omega_{c}  $. As such, in terms of the emitted photon frequency, the recoil acceleration is given by,
\bqe
a = \m \frac{\omega}{2\gamma_{i}}.
\eqe
Note, this is the same recoil acceleration which came from conservation of energy and momentum applied to a hyperbolic recoil, Eqn. (\ref{acc}), provided we identify the fractional energy loss $\m = \frac{13}{72}$. Thus, we have a qualitative agreement between the recoil acceleration produced via the LL equation and the assumption of an Unruh/hyperbolic recoil based on conservation of energy and momentum as well as the asymptotic analysis. We should note that using a value $\m = \frac{13}{17}$, then implies a $\chi = \frac{3 \mu}{2 \alpha} = 37$ which is a slight over estimation as we expect $\chi \leq 1.4$ \cite{2018NatCo...9..795W}. The difference between these values is likely due to the errors inherent in the asymptotic analysis. However, we are led to a rather intriguing expression for the recoil acceleration from radiation reaction,
\bqe
a \sim  \frac{\alpha \chi}{\gamma} \omega. \label{rraccel}
\eqe
This recoil acceleration is the result of the combined insight derived from experimental data, asymptotic time scale analyses, conservation of energy/momentum, the LL equation, and hyperbolic recoil.

\textbf{\underline{Power spectra below threshold:}} We have seen evidence that above the Unruh threshold, $\omega_c = \frac{E_i}{7}$ that radiation emission from the hyperbolic recoil, i.e. the Unruh effect, saturates the power spectra. However, below this threshold, we have a considerable departure of the theory from the data. In particular, we find an over estimation of the data for all data sets for $\omega < \omega_c$. Put differently, just as there appears to be an enhancement due to the Unruh effect in the high energy tail, there appears to be a depletion in the low energy portion. In the power spectra computed for the full LL trajectory \cite{1987JPhA...20.2405L, 1987A&A...176L..21L}, we do indeed see that the inclusion of radiation reaction yields a suppressed low energy portion. In fact, in our analysis we have a fractional energy loss of $\m = \frac{13}{72} = .18$ and in \cite{1987JPhA...20.2405L, 1987A&A...176L..21L}, a value of $\m = .2$ was specifically examined, and it was found to have nearly an order of magnitude depletion in the low energy portion, $\frac{\omega}{E_i} \sim .1$. The implication being that for the inclusion of radiation reaction in a high energy synchrotron power spectrum, the high energy portion will have an enhancement that is dominated by the Unruh effect while the low energy portion will see a depletion; both being the result of radiation reaction as described by the LL equation.

Finally, we note that the power spectra presented in \cite{1987JPhA...20.2405L, 1987A&A...176L..21L} was computed via the same $\frac{1}{\gamma}$ expansion as we employed here. However, in our analysis we specifically matched the series expansion in time, Eqn. (\ref{dv}) and (\ref{dz}), to that of the standard quasi-classical formalism \cite{1998ephe.book.....B} so we would have the same Airy function power spectrum for a direct comparison between theories. In \cite{1987JPhA...20.2405L, 1987A&A...176L..21L}, the full time series development, with recoil, is employed which results in a power spectrum which can only be computed numerically. Thus, in our Unruh analysis, we captured the salient features of radiation reaction in the energy high tail while the lower energy features require the full LL theory \cite{1987JPhA...20.2405L, 1987A&A...176L..21L}. What is of importance to note is the fact that we have now found the presence of the Unruh effect in the high frequency tail of LL radiation reaction.

\subsection{Frequency dependent temperature} \label{acce}

One of the more intriguing aspects of these ultra relativistic systems is the nature of the acceleration and thus temperature. In particular, how the temperature manifests in the data is itself noteworthy. One of the more simple examples is to consider a uniformly accelerated Unruh-DeWitt detector. If we consider an ensemble of detectors, we expect the ratio of excited to de-excited detectors to obey detailed balance, $\Gamma_{\uparrow} = \Gamma_{\downarrow}e^{-\Delta E/T_{FDU}}$. However, the fact that we need each detector to have both the same acceleration and energy gap has been prohibitively difficult for this pathway to be realized experimentally. Fortunately, radiative systems in particle physics have provided a resounding success. These observations of the Fulling-Davies-Unruh effect, measured by a photon spectrum, where the acceleration/temperature depends explicitly on the kinetic energy of an accelerated electron/positron \cite{lynch2019accelerated, 2024PTEP.2024b3D01L, 2024PhRvD.109j5009L, 2025GReGr..57..116L}. What is intriguing however, is the manner in which this temperature presents itself in the measured data.

In the case of the 1-dimensional thermal photon spectrum emitted via the radiative decay mode of the free neutron measured by the NIST-RDK-II collaboration \cite{nico2006observation, bales2016precision}, it was found that, under electron-mirror duality, the temperature of the system is set by the available budget of kinetic energy of the emitted electron \cite{2023FoPh...53...53G, 2024PTEP.2024b3D01L}, $T_{FDU} = \frac{6}{\pi^2}E_{kin}$. In turn, the kinetic energy of the electron is then set by the available energy budget from neutron decay, $E_{kin} \leq m_{n} - m_{p} - m_{e^-} - m_{\nu} = 782.3$ keV. However, this energy budget is also distributed into the kinetic energy of all daughter products of the neutron decay. Thus, the actual kinetic energy of the electron, and therefore temperature, is probabilistic \cite{1985JPhG...11..359V}. Surprisingly, it was found that the actual temperature which characterizes the experimentally measured thermal photon spectrum is constant and determined by the \textit{average kinetic energy} \cite{2024PTEP.2024b3D01L}. Thus, given a fully probabilistic photon temperature, a constant temperature is what presents itself experimentally in a measured photon spectrum.

In the CERN-NA63 experiments, a similar scenario also manifests. In the case of radiation reaction we expect an acceleration/temperature which is set by the recoil kinetic energy, $a \sim \frac{\omega^2}{m}$, of photon emission. Likewise, as in the case of radiative beta decay, a preliminary analysis, based on the standard form of Unruh-DeWitt detector energy gaps, found that a best fit constant acceleration also correctly describes the data \cite{lynch2021experimental}. Building upon this initial observation, we have now incorporated a hyperbolic recoil trajectory into the standard ultra relativistic synchrotron theory used for radiation reaction. With a threshold frequency of $\omega_c = \frac{E_i}{7}$ determined by the asymptotic analysis, we examined the accelerations produced by conservation of momentum both with and without conservation of energy, see section (\ref{econ}). Without conservation of energy, we reproduce the $a = \frac{\omega^2}{m}$ recoil acceleration identically. With conservation of energy, evaluated at $\omega_c$, we find a linear acceleration, $a = \frac{13 \omega}{144 \gamma_i}$, which successfully describes the experimental data above threshold. Without evaluating the acceleration at threshold, we have $a \sim \frac{\omega^2}{4 \gamma E_i}$. This is consistent with LL radiation reaction where an initial recoil acceleration is given by $a = \frac{2 \alpha \hbar \Omega_{c}^2 \gamma^2}{3 m c} = \frac{\alpha \omega^2}{6 \gamma E_{i}}$. Then, by substitution of the fractional energy loss invariant, $\m$, we have the linearly dependent $a = \m \frac{\omega}{2 \gamma_i}$ which matches the recoil acceleration based on conservation of energy. Thus, with either conservation of energy/momentum or with LL radiation reaction, the derived recoil acceleration which matches the data requires the \textit{linearized acceleration} by evaluation at threshold, $\omega_c$, or substitution of the invariant, $\m$, to match the data.

The above discussions detailing how the presence of acceleration/temperature manifest in data, e.g averages or linearization, apply to the measurement of photon spectra. The presence of the Unruh effect also implies the presence of an associated Rindler horizon which provides a dual thermodynamic description of the data. This is based on the second law of thermodynamics, $dQ = k_{B}T_{FDU}dS$, and its mapping to a particle spectrum \cite{2025GReGr..57..116L}. By identifying the heat flux through the horizon with the emitted photons, $dQ \sim d \omega$, and mapping the residual entropy in thermal radiation to photon number \cite{2016PhLB..757..383A}, $dS \sim dN$, We arrive at the photon spectrum, $\frac{dN}{d\omega} \sim \frac{1}{T_{FDU}}$. Thus, from a thermodynamic standpoint, the spectra measured at CERN-NA63 are, in fact, a direct measurement of the full frequency dependent FDU temperature, $T_{FDU} \sim \frac{\omega^2}{m}$ \cite{2025GReGr..57..116L}. This point serves to highlight the fact that not only can the signatures of the Unruh effect be measured via a particle spectrum of radiative photon emission \cite{lynch2021experimental, 2024PTEP.2024b3D01L}, but can also be directly probed via a thermodynamic/information analysis based on the second law of thermodynamics \cite{2025GReGr..57..116L}; the latter being a manifestation of particle creation via entropy, i.e. ``it from bit" \cite{2025arXiv250817067G}. 

\section{conclusions}
In this manuscript, we have analyzed a suite of high energy channeling radiation data sets measured by the CERN-NA63 collaboration. We extended the standard synchrotron theory to incorporate a hyperbolic recoil, and thus the Unruh effect, and applied it to the data sets. We find that above a critical frequency, defined by the radiation formation time, the dominance of the hyperbolic recoil over the standard synchrotron emission. The Unruh/recoil theory, with all parameters completely fixed by theory, provides a satisfactory description of the enhancement in the high frequency tail of the data.

The inclusion of the hyperbolic recoil into the trajectory provides a rather simple method to incorporate the Unruh effect into the standard prescriptions of strong field QED; namely ultra relativistic synchrotron emission. This further bolsters the analysis catalog of quantum radiation reaction which already includes techniques related to the quasi-classical formalism, stochastic models of radiation reaction \cite{2018NatCo...9..795W}, and the Landau-Lifshitz equation \cite{2021NJPh...23h5001N}. The presence of the Unruh effect in these systems, complimented by a relatively simple theoretical description based on the standard synchrotron theory, provides a novel baseline of analysis for future experiments in strong field QED. 
 
\section*{Acknowledgments}
The author wishes to thank Karen Hatsagortsyan for many stimulating discussions.

\bibliography{ref}

\end{document}